\documentclass[useAMS,usenatbib,usegraphicx]{mn2e}
\usepackage{natbib}
\usepackage{color}
\usepackage{mathrsfs}
\usepackage{times,xspace,amssymb}
\usepackage{url}
\usepackage{hyperref}
\usepackage{amsmath}
\bibpunct{(}{)}{;}{a}{}{,}
\def\rblr{R_{\rm BLR}}

\def\halpha{H$\alpha$\xspace}
\def\hbeta{H$\beta$\xspace}
\def\sii{[S\textsc{ ii}]\xspace}
\def\ha{H$\alpha$\xspace}
\def\hb{H$\beta$\xspace}
\def\lya{Ly$\alpha$\xspace}
\def\civ{C\textsc{ iv}\xspace}
\def\nv{N\textsc{ v}\xspace}
\def\nii{[N\textsc{ ii}]\xspace}
\def\heii{He\textsc{ ii}\xspace}
\def\feii{Fe\textsc{ ii}\xspace}

\def\rnlr{R_{\rm NLR}}

\def\zsun{Z_{\odot}}

\def\sunm{M_{\odot}}

\def\ergs{\ifmmode {\mathrm{ erg~ s^{-1}}} \else {\rm erg~s$^{-1}$}\ \fi}
\def\kms{\ifmmode {\mathrm{ km~ s^{-1}}} \else {\rm km~s$^{-1}$}\ \fi}

\def\mbh{M_{\bullet}}
\def\mmbh{m_{\bullet}}

\def\mgii{\ifmmode Mg {\sc ii} \else Mg {\sc ii}\ \fi}
\def\oiii{[O\textsc{ iii}]\xspace} %no space

\def\sunm{M_{\odot}}
\def\msun{M_{\odot}}

\newcommand{\dl}{$\lambda\lambda$}
\def\mnras{MNRAS}
\def\apj{ApJ}
\def\aj{AJ}

\def\aap{A\&A}

\def\apjs{ApJS}

\def\araa{Ann. Rev. A \& A}
\def\nat{Nature}

\def\pasp{Pub. Ast. Soc. Pacific}

\def\lesssim{\la}
\newcommand\phd{\phantom{.}}
\DeclareGraphicsExtensions{.ps}
\title[The BLR-NLR metallicity correlation]{%
Outflows from active galactic nuclei: The BLR-NLR metallicity correlation}
\author[Du et al.]{%
Pu Du$^{1}$\thanks{E-mail: dupu@ihep.ac.cn},
Jian-Min Wang$^{1,2}$,
Chen Hu$^{1}$,
David Valls-Gabaud$^{3,1,2}$,
Jack A. Baldwin$^{4}$,
\newauthor
Jun-Qiang Ge$^{2}$ and
Sui-Jian Xue$^{2}$%
\\
$^{1}$
Key Laboratory for Particle Astrophysics,
Institute of High Energy Physics,
 Chinese Academy of Sciences, 19B Yuquan Road,
Beijing 100049, China.\\
$^{2}$
National Astronomical Observatories of China,
Chinese Academy of Sciences,
 20A Datun Road, Beijing 100020, China.\\
$^{3}$
LERMA, CNRS UMR 8112, Observatoire de Paris,
61 Avenue de l'Observatoire,
 75014 Paris, France.\\
$^{4}$
Physics and Astronomy Department,
3270 Biomedical Physical Science Building,
 Michigan State University, East Lansing,
MI 48824, USA.
}
\begin{document}

\date{Accepted ... Received ... ; in original form ... }

\pagerange{\pageref{firstpage}--\pageref{lastpage}}
\pubyear{2013}

\maketitle

\label{firstpage}

\begin{abstract}
The metallicity of active galactic nuclei (AGNs), which can be measured by
emission line ratios in their broad and narrow line regions (BLRs and NLRs),
provides invaluable information about the physical connection between the
different components of AGNs. From the archival databases of the {\em
International Ultraviolet Explorer}, the {\em Hubble Space Telescope} and the
{\em Sloan Digital Sky Survey}, we have assembled the largest sample available of
AGNs which have adequate spectra in both the optical and ultraviolet bands to
measure the narrow line ratio \nii/\halpha and also, in the same objects, the
broad-line \nv/\civ ratio. These permit the measurement of the metallicities in
the NLRs and BLRs in the same objects. We find that neither the BLR nor the NLR
metallicity correlate with black hole masses or Eddington ratios, but there is
a strong correlation between NLR and BLR metallicities. This metallicity
correlation implies that outflows from BLRs carry metal-rich gas to NLRs at
characteristic radial distances of $\sim 1.0$ kiloparsec.  This
chemical connection provides evidence for a kinetic feedback of 
the outflows to their hosts.
Metals transported into the NLR enhance the cooling of the ISM in this region,
leading to local star formation after the AGNs turn to narrow line LINERs. This
post-AGN star formation is predicted to be observable as an excess continuum
emission from the host galaxies in the near infrared and ultraviolet, which
needs to be  further explored.
\end{abstract}
\begin{keywords}
   galaxies: nuclei
-- black hole physics
-- galaxies: active
-- galaxies: abundances
-- accretion
-- ISM: outflows
\end{keywords}

\section{Introduction}
Accreting supermassive black holes (SMBHs) are responsible for the major
phenomena observed in AGNs \citep[e.g., see
the recent review by][]{ho2008}, but also  strongly influence  their host
galaxies through kinetic/radiative
feedback processes \citep{dimatteo2005,bower2006,croton2006},
leading to a co-evolution between the SMBHs and
their hosts \citep[e.g.][]{magorrian1998,richstone1998}. Many details
of the specific feedback mechanisms remain under debate, but outflows have
been recognised to play
a key role in this co-evolution. Outflows transport energy and gas outwards from
the central regions, and also
carry metals to regions of galactic scales. The metallicity of the gas in AGNs
can be measured in both their broad
and narrow line regions, and might trace the physical relation between BLRs
and NLRs. This would provide a new
window for exploring the details of the feedback process.

The metallicity in broad line regions (BLRs) of AGNs has been extensively studied
following the
seminal paper of \cite{hamann1992}. Several line ratios have been suggested to
measure the BLR metallicity,
with the \nv/\civ intensity ratio being the most widely used.
Many studies have confirmed the following two basic properties of the BLR
metallicity: (1) it is very large, reaching
as much as 10 or more times the solar abundance
\citep{hamann1992,hamann1993,hamann2002,dhanda2007};
and (2) it has no cosmological evolution up to redshifts
$z=6.5\sim7.0$ \citep[e.g.][]{jiang2008,juarez2009,mortlock2012}. There are
additional --sometimes conflicting-- results suggesting that metallicity
correlates with either luminosity \citep{hamann1999,dietrich2002},
black hole masses \citep{warner2004,matsuoka2011a,matsuoka2011b} or Eddington
ratios \citep{shemmer2002,shemmer2004} for different samples. These properties
hint at a possible production
of metals in the nuclear regions and
is only a local event since its timescale is much shorter than
the local Hubble time. In this picture, black holes clean up the surrounding
metals before the next episode of activity.
This probably is linked with accretion flows, in which self-gravity drives
star formation \citep{collin1999,wang2010,wang2011,wang2012}.

It is far more difficult to use the \nv/\civ or similar ratios of UV lines to
measure the metallicity in NLRs in type I
AGNs, because the narrow components of lines such as \nv are very faint
\citep{nagao2006a}. Only about a
dozen high-redshift radio galaxies and type 2 AGNs have been measured for
metallicity in this way \citep{nagao2006a,matsuoka2009},
for which the NLR metallicity was found to lie in the range
$0.2 \zsun \lesssim Z \lesssim 5.0\zsun$ with
no cosmic evolution between redshifts $1.2\le z\le 3.8$. This is interesting
since it shows that the NLR metallicity has
properties similar to those of the BLRs. The NLRs have a size\footnote{
Note 
that, in our context, the NLR does not refer to the extended narrow line regions 
(ENLRs), which
are measured by \cite{greene2011}. The NLR size scaling law does not 
apply to  ENLRs.
\cite{fu2007} find that ENLRs only appear in objects with BLR metallicities
smaller than $\sim 0.6Z_{\odot}$. The typical size of NLRs of the
present sample is 0.7--1.0 kpc given
their \oiii\, luminosity of $\sim 10^{41}\ergs$, 
and correspond to the bulge regions 
\citep{simard2011}. The NLR and the bulge share the same approximate size. }
$R_{\rm NLR}=2.1~L_{\rm [O~III], 42}^{0.5}$ kpc, where
$L_{\rm [O~III],42}=L_{\rm [O~III]}/10^{42}{\rm erg~s^{-1}}$
is the \oiii$\lambda5007$ luminosity
\citep[e.g.][]{bennert2002,schmitt2003,netzer2004}. This is the typical size
of a bulge. It is thus expected that
the chemical properties of bulge gas can be deduced from the NLRs.
However, in these type 2 objects there are no broad
\nv or \civ lines to  measure  the BLR metallicity. Obviously what is needed are
measurements of both the BLR
and NLR metallicities in the {\it same} objects, meaning in type I AGNs.
This has never been done thus far.

In the interpretation of the classical Baldwin-Phillips-Terlevich
\citep[BPT][]{bpt1981} diagnostic diagram, it has been suggested that the
\nii/\halpha\ line ratio acts as an indicator of the metallicity of NLRs while
the \oiii/\hbeta ratio mainly depends on the ionization parameter
\citep[see Fig. 2 in][]{groves2006, storchi1989, storchi1991,
storchi1998, groves2004, stasinska2006, nagao2006c, kewley2013}.  This allows us to
measure the NLR metallicity for a given ionisation parameter in type 1 AGNs and
quantitatively compare it with the metallicities of the host galaxy and also of
the BLR. This line ratio also provides an opportunity to {\it simultaneously}
measure both BLR and NLR metallicities of type 1 AGNs in the same object.

In this paper we assemble from  published work and data
archives a sample of AGNs with spectroscopy covering
a wide wavelength range that includes \nv$\lambda$1240, \civ$\lambda$1550
, \nii$\lambda$6584, \hbeta and \halpha.
This sample consists of 31 type 1 AGNs, which
allow us to simultaneously measure the BLR and NLR metallicities, our goal being
to test the possible connection between
BLRs and NLRs. In \S2 we describe the sample of AGNs, and our measurements
of the BLR and NLR metallicities from both optical
and UV spectra. In \S3 we test for possible correlations between the NLR and
BLR metallicities and of these metallicities with either
black hole masses or Eddington ratios. The reliability and implications of
the correlations found are discussed in \S4. The
cosmological parameters $H_0=70 \,{\mathrm{ km~s^{-1}Mpc^{-1}}}$, $\Omega_m=0.3$
and $\Omega_{\Lambda}=0.7$ are assumed in this paper.

\section{The sample of AGNs with measured metallicities}
\label{section:2}

\subsection{Metallicity indicators}

Our aim is to compare the BLR and NLR metallicities on an object by object
basis.  We use the \nv/\civ ratio as a surrogate for the BLR metallicity
following the well-established techniques developed by
\cite{hamann1992,hamann1993,hamann1999}, who showed the ratio is
only sensitive to metallicity rather than temperature, ionization parameter and
gas density. Secondary production of N out of C and O via
CNO burning dominates at metallicities $Z\gtrsim0.2Z_{\odot}$, which result in
abundance of N $\propto Z^2$, making \nv/\civ flux ratio can be used as a 
metallicity indicator \citep{hamann1992,hamann1993,hamann1999,hamann2002}.
We do not use the \feii/H$\beta$ ratio as suggested by
\cite{netzer2007} since its possible correlation with BLR metallicity is a
secondary result based on comparing the correlation of different line ratios
with the Eddington ratio. The \feii excitation mechanism is poorly understood
and \cite{verner2004} found that the \feii strength is not sensitive to
metallicity.  

Metallicity in NLR can be indicated by the ratio of \nii/H$\alpha$ as shown
by Groves et al. (2006). In such a case, the ratio of narrow component \nv/\civ\, is very
tough to separate them from the spectra. The ratio of \nii/H$\alpha$ is
sensitive to metallicity when \nii/H$\alpha>0.1$ while \oiii/H$\beta$ is mostly sensitive
to the ionization parameter (Groves et al. 2006).

The considerations above indicate that we need a sample of
objects for which there are useful measurements of at least the
\nv$\lambda$1240, \civ$\lambda$1550, \hb, \oiii$\lambda$5007, \ha, and
\nii$\lambda$6584 lines.

\subsection{The samples of AGNs and quasars}
%\subsubsection{The optical/UV sample}
We have assembled a sample which includes all of the AGNs and quasars that we
know of for which it has been
possible to measure both the broad \nv and \civ emission lines at ultraviolet
wavelengths and the narrow
\hb, \oiii$\lambda$5007, \ha, and \nii$\lambda$6584 emission lines at optical
wavelengths. While there are about
900 AGNs in the \textsl{IUE} (International Ultraviolet Explorer) and
\textsl{HST} (Hubble Space Telescope)
data archives, only about 200 of these  have also optical spectroscopy
from the SDSS archive and published work.
Of this subset we select only Seyfert 1 galaxies and type 1 quasars, since
otherwise the \nv and \civ lines are obscured by dusty tori along the
line of sight. We  use either \textsl{HST} spectra directly or stack all
available \textsl{IUE} SWP images to produce
a single co-added \textsl{IUE} spectrum. Where possible we
use {\em HST} rather than {\em IUE} spectra to measure the \nv/\civ
ratios. We exclude objects with low signal-to-noise ratios in their
stacked UV spectra, broad absorption
line quasars and objects with strong \civ and \nv absorption. We  also
exclude objects having broad  Balmer
emission lines which are so strong that narrow emission lines
(i.e. \nii$\lambda6584$ and/or the narrow components
of H$\alpha$ and H$\beta$) cannot be measured, such as bright quasars
with ``disappearing'' narrow line regions
\citep{netzer2004}. The final sample contains 31 objects for which we can
measure both the BLR metallicity and the NLR
line ratios.
Table \ref{table1} provides the  details of the observations used
in the sample.

\begin{table*}
\begin{minipage}{130mm}
\caption{Optical and UV spectra of the AGN sample.}
\label{table1}
\begin{tabular}{lllccc}
\hline
\multicolumn{1}{c}{Object} &
\multicolumn{2}{c}{Sources and references for the spectra} &
$z$ &
$\log L_{5100}$ &
\multicolumn{1}{c}{$R_{\rm BLR}$} \\ \cline{2-3}
                               &
\multicolumn{1}{c}{Optical}     &
\multicolumn{1}{c}{Ultraviolet} &
                               &
(erg~s$^{-1}$)                  &
(light-day)           \\
\multicolumn{1}{c}{(1)}         &
\multicolumn{1}{c}{(2)}         &
\multicolumn{1}{c}{(3)}         &
\multicolumn{1}{c}{(4)}         &
\multicolumn{1}{c}{(5)}         &
\multicolumn{1}{c}{(6)} \\
\hline
2E 1615+0611                       & SDSS            & IUE (SWP39443, 44105, 55613)    & 0.038 & 43.25          &20.2\phd                      \\
Fairall 9{$^{\ast}$}    & ESO 1.5m (1)    & HST/FOS (2)                     & 0.047 & 44.25          &16.3$^{\dag}$                 \\
HB89 1028+313                      & SDSS            & IUE (SWP28214)                  & 0.178 & 44.45          &58.8\phd                      \\
Mrk 42                             & SDSS            & IUE (4 spectra stacked)         & 0.025 & 42.88          &\phd\phd 9.0\phd              \\
Mrk 106                            & SDSS            & IUE (SWP26911)                  & 0.123 & 44.52          &64.0\phd                      \\
Mrk 110                            & SDSS            & IUE (SWP33002)                  & 0.035 & 43.02          &18.8$^{\dag}$                 \\
Mrk 142                            & SDSS            & IUE (SWP20113, 20121)           & 0.045 & 43.61          &21.6\phd                      \\
Mrk 290                            & SDSS            & IUE (5 spectra stacked)         & 0.030 & 43.60          &21.3\phd                      \\
Mrk 359                            & ESO 1.5m (3)    & IUE (4 spectra stacked)         & 0.017 & 43.20          &13.3\phd                      \\
Mrk 493                            & OHP 1.93m (4)   & HST/FOS (5)                     & 0.031 & 43.27          &14.4\phd                      \\
Mrk 618{$^{\ast}$}      & WHT 4.2m (6)    & IUE (SWP40871)                  & 0.035 & 43.60          &21.3\phd                      \\
Mrk 771{$^{\ast}$}      & SDSS            & IUE (SWP16880, 16884, 20142)    & 0.063 & 43.92          &50.0$^{\dag}$                 \\
Mrk 841                            & McD 2.1m (7)    & IUE (17 stacked spectra)        & 0.036 & 43.84          &28.0\phd                      \\
Mrk 876                            & CA 2.2/3.5m (8) & IUE (11 stacked spectra)        & 0.129 & 45.00          &39.0$^{\dag}$                 \\
Mrk 1018{$^{\ast}$}     & SDSS            & IUE (6 stacked spectra)         & 0.042 & 43.85          &28.7\phd                      \\
Mrk 1126                           & ESO 1.5m (3)    & IUE (SWP36653, 36659)           & 0.011 & 42.71          &\phd\phd 7.4\phd              \\
Mrk 1239                           & OHP 1.93m (4)   & IUE (6 stacked spectra)         & 0.020 & 43.30          &14.9\phd                      \\
Mrk 1243                           & SDSS            & IUE (SWP36410)                  & 0.035 & 43.37          &16.1\phd                      \\
Mrk 1514                           & WHT 4.2m (6)    & HST/FOS (5)                     & 0.016 & 43.41          &\phd\phd 4.9$^{\dag}$         \\
NGC 3783{$^{\ast}$}     & ESO 1.5m (9)    & HST/FOS (2)                     & 0.010 & 43.12          &\phd\phd 4.5$^{\dag}$         \\
NGC 4051                           & Bok 2.3m (10)   & IUE (31 stacked spectra)        & 0.002 & 41.78$^{\dag}$ &\phd\phd 6.5$^{\dag}$         \\
NGC 4593                           & WHT 4.2m (11)   & IUE (29 stacked spectra)        & 0.009 & 42.75$^{\S}$   &\phd\phd\phd 1.2$^{\dag\dag}$ \\
NGC 5548                           & SDSS            & HST/FOS (2)                     & 0.017 & 43.31$^{\S\S}$ &21.2$^{\dag}$                 \\
NGC 5940{$^{\ast}$}     & SDSS            & IUE (SWP20821, 20832)           & 0.034 & 43.36          &15.8\phd                      \\
PG 0906+484{$^{\ast}$}  & SDSS            & IUE (SWP21921, 32463)           & 0.117 & 44.10          &38.7\phd                      \\
PG 0923+129                        & SDSS            & IUE (SWP25826)                  & 0.029 & 43.39          &16.5\phd                      \\
PG 1049-005                        & SDSS            & HST/FOS (2)                     & 0.360 & 45.58          &227.6\phd\phd\phd             \\
PG 1202+281                        & SDSS            & HST/FOS (2)                     & 0.165 & 44.21          &44.3\phd                      \\
SBS 1150+497                       & SDSS            & HST/FOS (5)                     & 0.334 & 44.71          &80.1\phd                      \\
Ton 256                            & SDSS            & IUE (SWP10071)                  & 0.131 & 44.52          &64.2\phd                      \\
US 1329                            & SDSS            & IUE (SWP56263)                  & 0.254 & 45.29          &161.1\phd\phd\phd             \\
\hline
\end{tabular}

\medskip

Columns (1), (2) and (3) list the AGN name and the source of spectral
data in the optical and UV bands, respectively; columns (4),
(5) and (6) give the redshift, the continuum luminosity at 5100 \AA\,
and  the BLR size.
For most objects, $L_{5100}$  is measured from the optical spectra,
except the following ones, for which it was taken from these references:
$^\dag$ direct reverberation mapping by \cite{kaspi2000},
$^{\dag\dag}$ from \cite{peterson2004},
$^\S$ from \cite{smith2004}, and $^{\S\S}$ from \cite{bentz2009}.
$^{\ast}$ are the sources whose broad \ha\ and \hb\ components have
different shapes.
References: (Columns 2 and 3): (1) \cite{santos1997}; (2) \cite{kuraszkiewicz2002};
(3) \cite{giannuzzo1996};
(4) \cite{veron2001}; (5) \cite{kuraszkiewicz2004};
(6) \cite{mullaney2008};
(7) \cite{grupe2004}; (8) \cite{kollatschny2006}; (9) \cite{stirpe1994};
(10) \cite{moustakas2006}; (11) \cite{dietrich1994}. 
Spectral resolutions: IUE: $\lambda/\Delta\lambda\sim 250$; 
HST: $\lambda/\Delta\lambda\sim 1300$ in UV bands; SDSS: $\lambda/\Delta\lambda\sim 1800$;
while the other instruments of telescopes listed in the table have $\lambda/\Delta\lambda\gtrsim 2000$.
\end{minipage}
\end{table*}

\begin{table*}
\begin{minipage}{130mm}
\caption{Measured and inferred quantities.}
\label{table2}
\begin{tabular}{lrrrrcc}
\hline
\multicolumn{1}{c}{Object} &
\multicolumn{1}{c}{$\log \left( \frac{%
\displaystyle\mathrm{[N\textsc{ii}]}}
{\displaystyle\mathrm{H\alpha}}\right)$} &
\multicolumn{1}{c}{$\log \left(\frac{%
\displaystyle\mathrm{[O\textsc{iii}]}}
{\displaystyle\mathrm{H\beta}}\right)$} &
%\multicolumn{1}{c}{$\log \left(\frac{%
% \displaystyle\mathrm{[N\textsc{ii}]}}
%{\displaystyle\mathrm{[O\textsc{iii}]}}\right)$}  &
\multicolumn{1}{c}{$\log \left(\frac{%
\displaystyle\mathrm{N\textsc{v}}}
{\displaystyle\mathrm{C\textsc{iv}}}\right)$} &
\multicolumn{1}{c}{FWHM} &
\multicolumn{1}{c}{$\log \mbh$} &
\multicolumn{1}{c}{$\log L_{\mathrm{Bol}}$} \\
 &
 &
 &
 &
$\left( \mathrm{ km~s^{-1}} \right)$ &
$\left( M_{\odot} \right)$ &
$\left( \mathrm{ erg~s^{-1}} \right)$ \\
\multicolumn{1}{c}{(1)} &
\multicolumn{1}{c}{(2)} &
\multicolumn{1}{c}{(3)} &
\multicolumn{1}{c}{(4)} &
\multicolumn{1}{c}{(5)} &
\multicolumn{1}{c}{(6)} &
\multicolumn{1}{c}{(7)} \\
\hline
       2E 1615+0611 &  $-0.44\pm0.01$ & $0.97\pm0.01$ & $-1.00\pm0.07$ &  3178 &  7.59 &  44.22 \\
  Fairall 9$^{\ast}$&  $-0.17\pm0.07$ & $0.95\pm0.04$ & $-0.49\pm0.01$ &  5858 &  8.17 &  45.12 \\
      HB89 1028+313 &  $-0.35\pm0.03$ & $1.04\pm0.02$ & $-0.57\pm0.13$ &  5794 &  8.73 &  45.31 \\
             Mrk 42 &  $-0.35\pm0.01$ & $0.10\pm0.02$ & $-0.73\pm0.09$ &   973 &  6.37 &  43.90 \\
            Mrk 106 &  $-0.40\pm0.02$ & $0.95\pm0.03$ & $-0.82\pm0.28$ &  4108 &  8.47 &  45.38 \\
            Mrk 110 &  $-0.79\pm0.02$ & $1.01\pm0.01$ & $-1.08\pm0.12$ &  2356 &  7.02 &  44.02 \\
            Mrk 142 &  $-0.47\pm0.03$ & $0.48\pm0.13$ & $-0.48\pm0.10$ &  1773 &  7.27 &  44.54 \\
            Mrk 290 &  $-0.45\pm0.03$ & $1.01\pm0.03$ & $-0.64\pm0.03$ &  4000 &  7.97 &  44.53 \\
            Mrk 359 &  $-0.41\pm0.01$ & $0.62\pm0.02$ & $-0.89\pm0.03$ &   887 &  6.46 &  44.18 \\
            Mrk 493 &  $-0.29\pm0.01$ & $0.12\pm0.01$ & $-0.37\pm0.01$ &   886 &  6.49 &  44.24 \\
    Mrk 618$^{\ast}$&  $ 0.02\pm0.01$ & $0.72\pm0.03$ & $-0.32\pm0.05$ &  2356 &  7.51 &  44.53 \\
    Mrk 771$^{\ast}$&  $ 0.13\pm0.05$ & $1.14\pm0.04$ & $-0.36\pm0.02$ &  3048 &  8.15 &  44.82 \\
            Mrk 841 &  $-0.39\pm0.03$ & $1.10\pm0.03$ & $-0.90\pm0.33$ &  4085 &  8.11 &  44.75 \\
            Mrk 876 &  $-0.16\pm0.04$ & $0.64\pm0.08$ & $-0.74\pm0.08$ &  7781 &  8.65 &  45.83 \\
   Mrk 1018$^{\ast}$&  $ 0.22\pm0.06$ & $0.77\pm0.07$ & $-0.22\pm0.20$ &  4366 &  8.18 &  44.76 \\
           Mrk 1126 &  $ 0.09\pm0.01$ & $0.98\pm0.02$ & $-0.57\pm0.19$ &  4882 &  7.68 &  43.75 \\
           Mrk 1239 &  $-0.30\pm0.01$ & $0.83\pm0.01$ & $-0.45\pm0.24$ &   951 &  6.57 &  44.26 \\
           Mrk 1243 &  $-0.45\pm0.05$ & $0.58\pm0.05$ & $-0.26\pm0.04$ &  1967 &  7.23 &  44.32 \\
           Mrk 1514 &  $-0.17\pm0.01$ & $0.43\pm0.01$ & $-0.76\pm0.01$ &  1967 &  7.08 &  44.36 \\
   NGC 3783$^{\ast}$&  $-0.08\pm0.03$ & $1.00\pm0.05$ & $-0.65\pm0.01$ &  2594 &  7.24 &  44.10 \\
           NGC 4051 &  $-0.57\pm0.03$ & $0.43\pm0.06$ & $-0.57\pm0.03$ &  1794 &  6.38 &  42.95 \\
           NGC 4593 &  $ 0.07\pm0.01$ & $1.06\pm0.03$ & $-0.45\pm0.06$ &  3826 &  6.67 &  43.78 \\
           NGC 5548 &  $-0.26\pm0.01$ & $1.02\pm0.01$ & $-0.97\pm0.02$ &  6378 &  8.36 &  44.28 \\
   NGC 5940$^{\ast}$&  $-0.11\pm0.03$ & $0.89\pm0.04$ & $-0.46\pm0.05$ &  3805 &  7.80 &  44.31 \\
PG 0906+484$^{\ast}$&  $-0.25\pm0.06$ & $0.98\pm0.08$ & $-0.46\pm0.12$ &  4691 &  8.37 &  44.99 \\
        PG 0923+129 &  $-0.32\pm0.02$ & $0.95\pm0.02$ & $-0.61\pm0.05$ &  2140 &  7.32 &  44.34 \\
        PG 1049-005 &  $-0.53\pm0.03$ & $1.03\pm0.02$ & $-0.90\pm0.05$ &  5513 &  9.28 &  46.38 \\
        PG 1202+281 &  $-0.53\pm0.02$ & $0.84\pm0.01$ & $-0.76\pm0.01$ &  4540 &  8.40 &  45.09 \\
       SBS 1150+497 &  $ 0.05\pm0.04$ & $1.23\pm0.06$ & $-0.77\pm0.05$ &  4064 &  8.56 &  45.55 \\
            Ton 256 &  $-0.51\pm0.02$ & $1.04\pm0.01$ & $-0.81\pm0.08$ &  2789 &  8.14 &  45.38 \\
            US 1329 &  $-0.98\pm0.06$ & $0.92\pm0.02$ & $-0.98\pm0.19$ &  3827 &  8.81 &  46.10 \\
\hline
\end{tabular}

\medskip

Col. (1): object; (2): \nii$\lambda6584$/H$\alpha$ flux ratio;
(3): \oiii$\lambda5007$/H$\beta$
flux ratio; (4): \nv$\lambda1240$/\civ$\lambda1550$ flux ratio;
(5): FWHM of H$\beta$; (6): SMBH mass; (7): bolometric luminosity.
$^{\ast}$ are  sources whose broad \ha\ and \hb\ components have
different shapes.
\end{minipage}
\end{table*}

\subsection{Spectral fitting}
\label{sec:fits}

Table \ref{table2} gives the measures of line ratios, \hb~FWHM, black hole
masses and bolometric luminosities of the selected AGNs. Following  \cite{hu2008},
we first correct the spectrum for Galactic extinction, shift it to
the galaxy's rest-frame, and fit a continuum model. The continuum model has
three components: (1) a single power law; (2) Balmer continuum emission; and (3)
a pseudo-continuum due to blended Fe emission.

Since the key point here is to get matched profile flux ratios to make sure
that the emission lines whose fluxes are compared come from the same region, we
adopt the following procedure to measure the emission lines after subtracting
the continuum. In the UV spectra we model the broad-emission
lines of \lya, \nv and \civ through two sets of Gaussian components each which
are forced to have the same FWHM and velocity shifts. The sum of the two
Gaussian components of each line is used to obtain fluxes of the line. 
\footnote{
The FWHM of the narrow component of NGC 5548's \nv and \civ emission lines 
is 1118 km/s, which is the
narrowest one in the present sample, and is marginally consistent with 
the criteria of broad emission line ($>$1000 km/s). But there is still possibility
that it comes from NLR, so we relaxed the constraint of tying the profiles
of the two gaussians used to fit \nv and \civ , and only use the broader components
to calculate NV/CIV ratio. In this case, $\log(\mathrm{\nv/\civ})$ of NGC 5548 
becomes $-0.80\pm0.04$. 
This number is still very similar to the value we used in the text, and does not 
change the conclusion at all.} In the
optical band, we fit the emission lines with three components, the same
component of each line forced to have the same profile: (1) a narrow component
modeled with one Gaussian each for \ha, \hb, \oiii\dl4959, 5007, \nii\dl6548,
6584, \sii\dl6717, 6731; (2) a broad component modeled with two Gaussians each
for \ha and \hb; and (3) a wing component modeled with one Gaussian for
\oiii\dl4959, 5007 when necessary. Among these lines, the flux ratios of
\oiii\dl4959, 5007, \nii$\lambda$6584 and \nii$\lambda$6548 are fixed to the
theoretical values of 3.0 and 2.96 respectively. There are seven sources
(listed in Tables \ref{table1} and \ref{table2}) showing different shapes of
the broad \ha and \hb components. We thus relax the constraint of the FWHM and
velocity shift of the corresponding Gaussians and add a third Gaussian component
to them. 
%These profiles are not constrained for seven objects
%(see Tables~\ref{table1} and \ref{table2}) whose broad
%\ha and \hb components clearly have different shapes, but this 
It does not significantly affect the correlations
among the line ratios and other quantities ($\mbh$, $L_{\rm Bol}/L_{\rm Edd}$).
We measure the ratios \oiii/\hb and \nii/\ha in the narrow components 
(not including the wing component of \oiii), 
and FWHM of broad component of \hb from the fitted gaussians which are used to model it.
The continuum-subtracted spectra are provided in the Appendix as well as the
corresponding results of emission-line fittings. The Appendix is provided on-line.

\subsection{Estimates of SMBH parameters}

To estimate the masses of the super-massive black holes (SMBHs)
in our UV/optical sample, we use the empirical
$R_\mathrm{BLR}-L$ relation, except in objects for which direct measures
from reverberation mapping observations
are available. Using the host-galaxy corrected relation
$R_{\rm BLR} = 34.4 \, L_{44}^{0.52}$ light days
\citep{bentz2009}, where $L_{44} = L_{5100}/10^{44} $ \ergs\,
and $L_{5100}$ is the continuum luminosity at 5100 \AA,
we have the SMBH mass from $\mbh = (f/G) R_\mathrm{BLR} V_\mathrm{H\beta}^2 $,
where $G$ is the gravitational constant,
$V_\mathrm{H\beta}$ is the full-width-half maximum of the
H$\beta$ emission line and $f$ is a factor determined
by the geometry of the BLR. We adopt $f=1.4$, following the calibration of
the relation linking the velocity dispersion
to the SMBH mass \citep{onken2004}. We obtain the bolometric luminosity
using an AGN continuum template
\citep{marconi2004}, and then the Eddington ratio from
$\mathscr{E}=L_\mathrm{Bol}/L_\mathrm{Edd}$, where
$L_\mathrm{Edd}=1.3 \times 10^{38} \, \mmbh $ \ergs\, is the Eddington limit
luminosity, $L_\mathrm{Bol}$ is the
bolometric luminosity and $\mmbh = \mbh / \msun$.

Note that the average contributions from host galaxies 
\citep{bentz2009} is about $1/2$ of the total 5100 \AA\, luminosity.
As $\rblr$ scales 
with the square root of the luminosity, the contaminations lead
to a further uncertainty of 0.15 dex to the 
estimations of black hole masses as well as to the Eddington ratios. We 
include them in their error budget. 
To be conservative, we set the error bars of black hole
 masses, Eddington ratios and bolometric
luminosity to 0.3 dex, as they are mainly determined by the 
systematic uncertainties of 
$f_{\rm BLR}$ in the $R_{\rm BLR}-L$ relation and host contaminations
\citep{onken2004, park2012, woo2010, woo2013}.

\begin{figure*}
\begin{minipage}{120mm}
\includegraphics[angle=0,width=120mm]{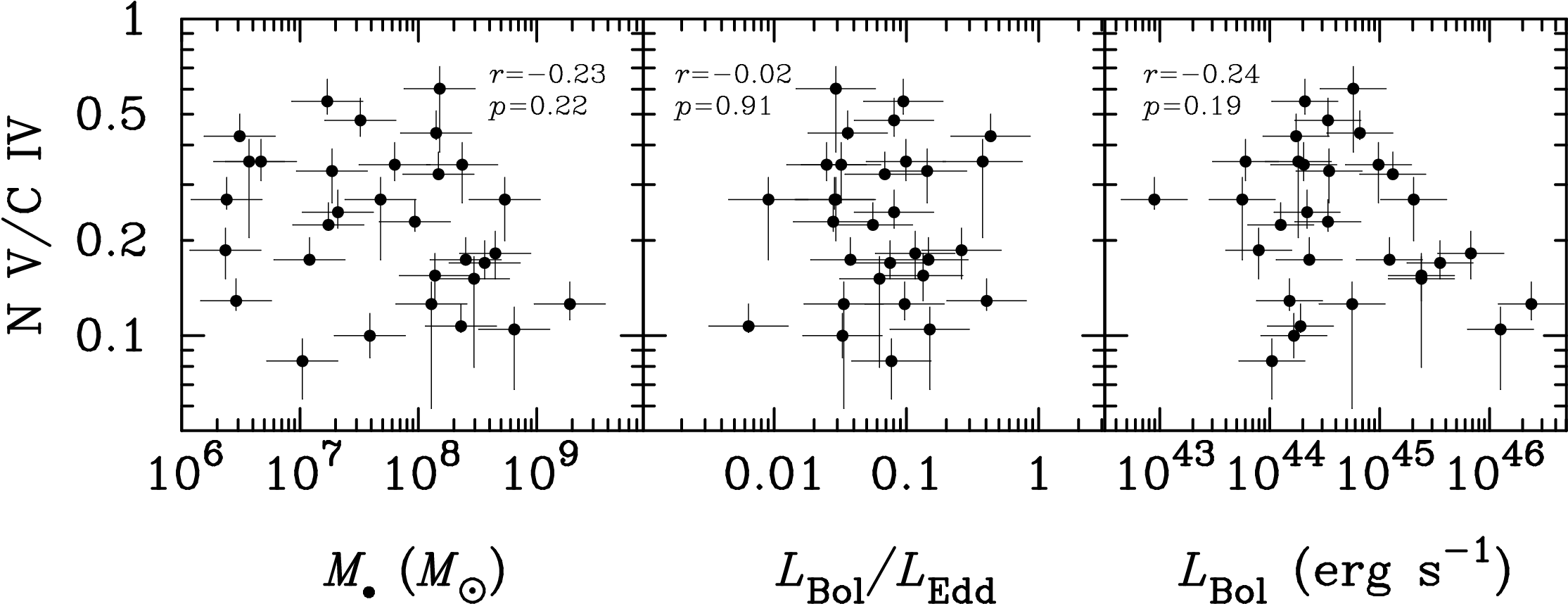}
\caption{ The line ratio of \nv/\civ\, (BLR metallicity) versus
black hole masses, Eddington ratios and bolometric luminosity.  The sample
shows no significant correlation between the metallicity and these parameters.
%We set the error bars of black hole masses, Eddington ratios and bolometric
%luminosity to be 0.3 dex, which are mainly determined by the uncertainties in
%$f_{\rm BLR}$ in the $R_{\rm BLR}-L$ relation and host contaminations.
Pearson's coefficient ($r$) and null probability ($p$) are given in each
panels. The BLR metallicity does not correlate with $\mbh$,
$L_{\rm Bol}/L_{\rm Edd}$ or $L_{\rm Bol}$.}
\label{fig1}
\end{minipage}
\end{figure*}

\begin{figure*}
\begin{minipage}{120mm}
\includegraphics[angle=0,width=120mm]{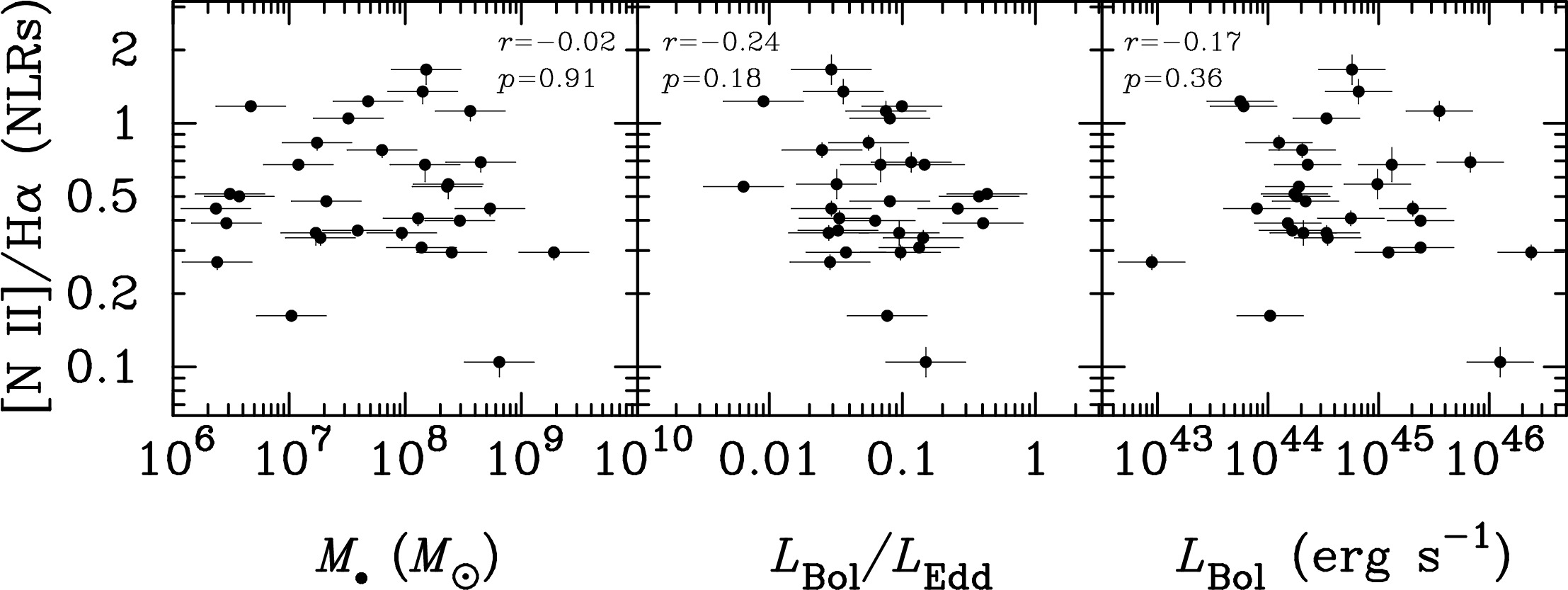}
\caption{The line ratio of \nii/H$\alpha$ (NLR metallicity)
versus black hole masses, Eddington ratios and bolometric luminosity.  
The sample shows no correlation
between the metallicity and these parameters. The error bars of black hole
masses and Eddington ratios are same as in Figure~\ref{fig1}.  Pearson's
coefficient and null probability are indicated in each panel. The
NLR metallicity does not correlate with any of these parameters.
}
\label{fig2}
\end{minipage}
\end{figure*}

\section{Analysis of correlations}
%---------------------------
\subsection{BLR metallicities}
%----------------------------
Figure~\ref{fig1} shows that, in our carefully-selected sample, the
BLR metallicities do {\it not}
correlate with either black hole masses 
or Eddington ratios, or bolometric luminosity. 
This result can be quantitatively established 
through the Pearson linear correlation coefficient ($r$) and 
associated probability($p$) that
$|r|$ be larger than the observed value if 
the null hypothesis of zero correlation is true. 
As indicated in Figure~\ref{fig1}, the
large $p$ values imply that the null hypothesis that
the variables are not correlated is accepted in all the cases
shown.
Previous
studies appear to give different results as listed in Table 3 for
a brief summary. \cite{warner2003}
compiled a list of 578 quasars (more than 800 spectra) with
redshifts $0<z<5$ to explore the potential relation of BLR
metallicity with black hole masses and Eddington ratios. They estimated black hole
masses through \civ profiles
for high$-z$ quasars, and then produced a series of composite spectra binned
by black hole mass. Their sample
spans 5 orders of magnitude in black hole masses and 6 orders in luminosity, and they
found a correlation between
$Z_\mathrm{BLR}$ and black hole masses from their stacked spectra. However,
\cite{shemmer2004} found from a
detailed analysis of high-quality spectra of 29 high-$z$ quasars that there
is {\it no} correlation between
$Z_{\mathrm{BLR}}$ and black hole masses, but instead a weak correlation
between the \nv/\civ ratios and the Eddington
ratios. Using stacked spectra of SDSS quasars with $2.3 < z < 3.0 $,
\cite{matsuoka2011a} found no
correlation between metallicity and Eddington ratios, but that one
does exist between $Z_\mathrm{BLR}$ and
black hole masses. Very recently, \cite{shin2013}
studied the metallicity of PG quasars
observed by {\em IUE} and {\em HST} and found that their metallicity does
not correlate with black hole masses, and perhaps
weakly with Eddington ratios. The inconsistencies between these results
might be due to selection effects or to
the way in which the spectra were combined into composite spectra in
some of the studies. The composite spectra employed in 
metallicity
analysis may be misled by the uncertainties in black hole mass. In particular,
the estimation of black hole 
masses in quasars through \civ\, should be improved somehow 
\citep{denney2013}, 
leading to large uncertainties of Eddington ratios.

We find that our results are consistent with those of \cite{shemmer2004}, in
spite of the fact that they found a weak correlation with Eddington ratio while
we do not.  This stems from the fact that \cite{shemmer2004} included some
high-redshift quasars and narrow-line Seyfert 1 galaxies, making their sample
to be  biased towards objects with larger Eddington ratios than ours. Except
for these objects, the weak correlation between $Z_\mathrm{BLR}$ and Eddington
ratios disappears. The correlation between $Z_{\rm BLR}$ and
Eddington ratios is only moderate in the upper right panel of Figure 8 
in \cite{shin2013}, and it disappears altogether when excluding 
objects with Eddington
ratios  $\le 10^{-2}$. Our conclusion is that previous studies do not
provide in fact any strong evidence for correlations of BLR metallicity with
black hole masses or Eddington ratios, in agreement with our findings.

Our work doesn't show the correlation of BLR metallicity
with bolometric luminosity which is common in previous works
\citep{shemmer2002, warner2004, nagao2006b, shin2013}. The reason is that the
bolometric luminosity of the present sample has a much smaller range than those
of the previous samples ($10^{44}-10^{45.5}~\mathrm{erg~s^{-1}}$ of $L_{\rm
Bol}$ in the present sample, but $10^{44.5}-10^{48.5}~\mathrm{erg~s^{-1}}$ of
$L_{\rm Bol}$ in \cite{warner2004}, $10^{10}-10^{14}~\mathrm{erg~s^{-1}}$ of
$\nu L_{\nu}(1450\mathrm{\AA})$ in \cite{shemmer2002} and $10^{44}-10^{46.8}~\mathrm{erg~s^{-1}}$
of $L_{\rm Bol}$ in \cite{shin2013}).  However, narrow spans of bolometric luminosity
and Eddington ratio may provide us opportunities to find correlations which are
easy to be concealed by large scatters of these parameters.

\begin{table*}
%\begin{minipage}{175mm}
\caption{A summary of the previous results of correlations of $Z_{\rm BLR}$ with central engines}
\label{table3}
\begin{tabular}{llllll}\hline
References   & sample     & $L_{\rm Bol}/L_{\rm Edd}$ & $\mbh$ & $L_{\rm UV}$ or $M_B$ & Notes\\ \hline
Hamann \& Ferland (1993)  & $\sim 100$ quasars ($0\lesssim z\lesssim 5$) & ...  & ... & yes, but weak & both \nv/\civ, \heii/\civ\\
Shemmer \& Netzer (2002)  & 162 AGNs (including 8 NLS1)                                    &...   &...  &yes &weakened by NLS1s \\
Warner et al. (2003; 2004)& 578 quasars ($0\lesssim z\lesssim 5$)        & No   &yes  & yes           & composite spectra\\
Shemmer et al. (2004)     & 29 quasars ($z\approx 2\sim 3$)+92 AGNs & ... & yes & very weak & only \nv/\civ\\
Nagao et al. (2006)       & $\sim 5000$ SDSS quasars DR2 ($2.0\lesssim z\lesssim 4.5$)&... & ... & yes & composite spectra \\
Matsuoka et al. (2011)    & $\sim 2383$ SDSS quasars  ($2.3\lesssim z\lesssim 3.0$)& yes & yes &yes &composite spectra, but only\\
                         &                                                        &     &     &    &\nv/\civ, \heii/\civ \\
Shin et al. (2013)        & 70 PG quasars ($z<0.5$)  &  yes, but moderate & very weak  &yes, but weak &both \nv/\civ, \heii/\civ \\ \hline
\end{tabular}
%\end{minipage}
\end{table*}

There are ten objects in common with the
sample of \cite{shin2013}, and we find that their measurements 
of \nv/\civ are in excellent agreement with ours, except for 
Mrk 771 and PG 1049-005.  In Mrk 771, \cite{shin2013} find  \nv/\civ$=0.10$, 
but the {\em IUE} spectrum we analysed shows a quite strong \nv
feature. In PG 1049-005, they find \nv/\civ$=0.71$ which 
may be possible, given 
the presence of strong wings in the \lya\ and \civ lines. A proper
check would require  high
quality UV spectra). These differences do not change our 
results.

\subsection{NLR metallicities}
%-----------------------------
The \nii/\halpha line ratio provides a
measure of the NLR metallicity. Figure~\ref{fig2} shows that
the NLR metallicity does not correlate with either
black hole masses, bolometric luminosity or Eddington ratios, 
as was the case for the BLR.

A typical NLR is about the same size as a bulge, and if its metallicity
is mainly contributed by
stellar evolution in the bulge then the two should have metallicities which
are roughly similar and which
would be expected to gradually increase with cosmic time. However, the NLR
metallicity in high-$z$
radio galaxies does not correlate with redshift
\citep{nagao2006a,matsuoka2009}. This
agrees with the present results that NLR metallicity neither correlates
with black hole masses nor Eddington ratios.

Additionally, the \cite{magorrian1998} relation (that black hole masses are linearly
proportional to bulges of hosts)
implies that if the NLR metallicity originates from stellar evolution of
bulges, there should be a correlation
between black hole mass and NLR metallicity due to the well-known correlation
between metallicity and galaxy masses
\citep{tremonti2004}. No such correlation is seen, again suggesting
that the NLR metallicity is not dominated by metal production from the host
galaxies. Furthermore, it has been found that the NLR metallicity has not undergone 
cosmic evolution \citep{matsuoka2009}, but there is a trend with luminosity like
the BLR metallicity \citep{nagao2006b}. However, the metallicity in star-forming galaxies
has significant cosmic evolution,  decreasing with redshift 
\citep[e.g.][]{shapley2005,erb2006}. 
%{\color{red}The average
%\nii/\halpha ratio of the present sample is significantly larger than
%that of star-forming galaxies.} 
The question that arises,
given that, the metallicity in NLRs maybe not produced by the host
galaxy, is: what could possibly be its origin?
%We note that the present sample is not large.

\subsection{The BLR-NLR metallicity correlation}
\begin{figure*}
\includegraphics[angle=0,width=\columnwidth]{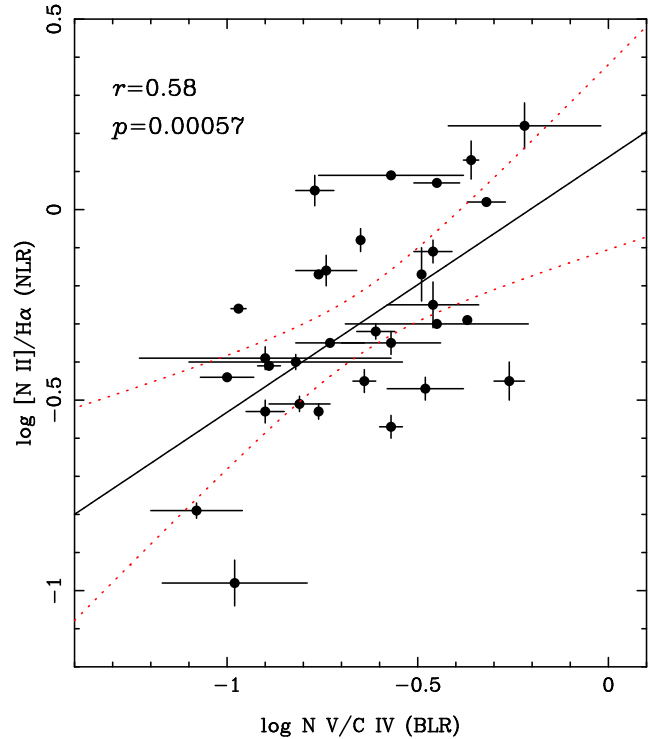}
\caption{NLR metallicities versus BLR metallicity. The correlation found
in this sample
is highly significant (the red dotted line mark the 95\%-level confidence
boundaries for the linear regression given by the black line) and
is not caused by ionisation parameters (see \S~\ref{sec:blrnlr} for details).}
\label{fig3}
\end{figure*}

%----------------------------------------------
\label{sec:blrnlr}
Figure~\ref{fig3} compares the NLR and BLR metallicities, and shows that
they are strongly correlated given the regression coefficient $r=0.58$
and the null probability of no correlation $p=5.7\times 10^{-4}$. The strong
correlation is surprising because BLRs and NLRs are separated by
scales of about 1.0 kiloparsec. Their metallicities
indicate that these two separated regions are in fact physically connected,
but {\it how} do they connect?

The correlation cannot be caused by ionisation parameter effects. The
\nv/\civ broad line ratio is
weakly proportional to the ionisation parameter ($\Xi_{\rm BLR}$)
\citep[see Fig. 4. in][]{hamann2002}
whereas the \nii/\ha\, narrow line ratio  is approximately inversely
proportional to $\log\Xi_{\rm NLR}$  (Fig.~
8 in \cite{dopita2000}, Fig.~8 in \cite{kewley2002} and Fig.~13 \cite{nagao2006c}).
On the other hand, $\Xi_{\rm BLR}$ and $\Xi_{\rm NLR}$ could be very different from 
each other. The present correlations
as shown by Fig.~\ref{fig3} are thus not be caused by $\Xi_{\rm BLR}$ and
$\Xi_{\rm NLR}$ (estimated in \S~4.2) in the present sample.

The spectral energy distributions (SEDs) of the central
 engines may influence the line
ratios of \nv/\civ. As the SEDs are  
controlled by the Eddington ratios
\citep{wang2004, shemmer2006, brightman2013},
the correlation between
BLR--NLR metallicity can not be caused by the different SEDs 
because the distribution of the Eddington ratios in the present 
sample is very narrow.
%we should note that the distribution of the Eddington
%ratios in the present sample is very narrow, therefore, the correlation between
%BLR-NLR metallicity can not be caused by the different SEDs of central engines.
The gas density in NLR should not exceed about $10^5{\rm cm^{-3}}$, above which
forbidden lines are highly suppressed. Furthermore, as indicators of
metallicity in BLR and NLR, the two line ratios are not sensitive to gas
density. 

We are therefore led to conclude that there is
a link between the metallicities in the NLR and in the BLR in
these objects.

\section{Discussion}
Several models exist in which BLR metals are produced by bursts of star
formation in the accretion flows that are
fueling the central black holes
\citep[e.g.]{collin1999,collin2008,wang2010,wang2011,wang2012}, but the
large metallicity in NLRs remains intriguing. The key question is to
understand how  the metallicity can be correlated in these two regions which
have such different characteristic sizes? Even though the 
ionised gas in NLRs is mainly
supplied by  stellar winds from the bulge, the NLR metallicity of the 
ionised gas
can be strongly affected by outflows developed from the central regions 
of AGNs. The 
metals delivered to the NLRs by the outflows, specifically by outflows 
from the
BLRs, might enhance the metallicity in NLRs.

\subsection{Metals produced in BLRs}
It is generally accepted that black holes grow episodically  with
cosmic time \citep{small1992,marconi2004,wang2006,wang2008}. The absence
of metallicity evolution on cosmological
timescales \citep{hamann1999,warner2003}
strongly indicates that, for every episode, the star forming discs
initially start from a larger
reservoir of gas having the metallicity of the host galaxy, rather than
being steadily built up
as the same gas cycles through many episodes. This is consistent with the
theoretical expectation
for a single episode of SMBH activity \citep{wang2010,wang2011}. In this model,
metals are produced in
the self-gravitating part of accretion
discs as a result of supporting the Shakura-Sunyaev disc \citep{shakura1973} around
SMBHs \citep{wang2010}.
BLRs are much more metal-rich than their host galaxies, and
therefore the excess
metals cannot be the products
of the stars in the host galaxies.  In this picture, a black hole's
accretion rate
stays approximately constant during a single AGN episode, but the BLR's
metallicity increases monotonically
with time, and is regarded as a probe of age of the particular AGN episode.
In this model,
no strong correlation is expected between metallicity and either the
black hole masses or Eddington ratios.

On the other hand, the peak metallicity reached during each episode
(Fig. 3 {\em left} panel in
\cite{wang2011}) might correlate with the Eddington
ratio. The individual objects within a sample of AGNs will have different
ages within their respective episodes,
and different samples will have different mixtures of ages, which may
explain the variety of results (i.e.
lack or presence of correlations between
metallicities and Eddington ratios, black
hole masses and luminosities) found in
different studies.

\subsection{Emitting clouds in NLRs}
Before addressing possible mechanisms which can account for the
observed correlation  between the metallicities in the BLR and NLR, the
issue of the role of variations in the ionisation parameter $\Xi$ must be
assessed in the robustness of the measure of metallicity through the
\nii/\halpha  line ratios.  Inspired
by the fact that the extended
NLRs are at a radial distance from the AGN that is similar to the size of
the bulge \citep{bennert2002,schmitt2003,netzer2004}, we explore the NLR
physics within
the framework of the coevolution of
SMBHs and galaxies \citep{richstone1998} in order to understand the
evolutionary
tracks of AGNs. A key assumption is that the NLR clouds are pressure
confined by
the hot gas from stellar winds in the bulge. The main features of the
NLR model
are: (1) a two-phase medium composed of cold clouds and a hot medium;
(2) a hot medium from stellar winds in the bulges \citep{ho2009}, ionised
by the central engine radiation
and cooled via line emission; (3) cold clouds formed from AGN
outflows whose evidence comes by the usual blue shift of \oiii; and
(4) a star forming
disc that  produces metals continuously, which are then transported
outwards by  AGN
outflows. The last two assumptions are supported by the strong correlation
of NLR
and BLR metallicities. Assuming photoionisation, we give a simple estimate
of the properties
of the NLR line emission as follows.

The ionisation parameter is defined by
$\Xi_{\rm NLR} \; = \; L_{\rm ion}/4\pi \, R_{\rm NLR}^2 \, c \,
n_e \, k \, T_e$,
where $L_{\rm ion}$ is the ionising luminosity, $c$ and $k$ are the speed
of light and the
Boltzmann constant, respectively, $R_{\rm NLR}$ is the NLR size, and $n_e$
and $T_e$ are the
density and the temperature of the ionised gas of cold clouds in the NLR.
Considering the relation
$\rnlr \, \propto \, L_{\rm [OIII]}^{1/2}\propto L_{\rm ion}^{1/2}$,
we have $\Xi_{\rm NLR} \, \propto \, (n_e T_e)^{-1}$.
We assume that the NLR is composed of numerous discrete clouds,
which are confined by the surrounding hot interstellar medium (ISM) from
stellar winds in
bulges \citep{ho2009}. In such a simple two-phase model, the pressure balance
equation reads $n_e T_e = n_h T_h$,
where $n_h$ and $T_h$ are the density and temperature of the hot ISM. For the
simplest consideration, the hot
ISM has a virialised temperature of
$kT_h\approx GM_{\rm bulge}/R_{\rm NLR}$. Considering the
Magorrian relation $\mbh\propto M_{\rm bulge}$ \citep{magorrian1998} and
$R_{\rm NLR}\propto L^{1/2}\propto \left(\mathscr{E}\mbh\right)^{1/2}$,
we have $T_h\propto \left(\mbh/\mathscr{E}\right)^{1/2}$.
It has been suggested that, in the bulge, the hot ISM is eventually accreted
onto the SMBH without going
through a static state \citep{ho2009}. For a simple version of the inflows
in the bulge, the mass rate is
given by
$\dot{M}_{\rm inflow} \, = \, 4\pi \, \rnlr^2 \, V_{\rm ff} \, n_h \, m_p$,
where $V_{\rm ff}=\left(GM_{\rm bulge}/\rnlr\right)^{1/2}$ is the free-fall
speed of the hot gas and
$M_{\rm bulge}$ is the mass of the bulge. We assume a simple relation
between the
SMBH accretion rate ($\dot{M}_{\bullet}$)  and the inflow, so that
$\dot{M}_{\bullet} \, \propto \, \mathscr{E}\mbh \, \propto \,
\dot{M}_{\rm inflow}$,
yielding the scaling relation
\begin{equation}
n_h \; \propto \; \mathscr{E} \; R_{\mathrm{NLR}}^{-3/2} \; M_{\rm bulge}^{1/2} \;
\propto \; \left(\frac{\mathscr{E}}{\mbh}\right)^{1/4} \; .
\end{equation}
As the thermal pressure of the hot ISM is
$n_hT_h \, \propto \, \left(\mathscr{E}/\mbh\right)^{-1/4}$, we find
that
\begin{equation}
\Xi_{\rm NLR}\; \propto \; \left(n_eT_e\right)^{-1} \; \propto \;
\left(\frac{\mathscr{E}}{\mbh}\right)^{1/4} \; .
\end{equation}
This yields a range of ionisation parameter as
$\Delta \log\Xi_{\rm NLR}=0.25\Delta\log\left(\mathscr{E}/\mbh\right)$. From
Fig.~\ref{fig1},
the scatter of the present sample is $\Delta\log \mathscr{E}\sim 2.0$
and $\Delta\log\mbh\sim 3.0$,
which implies that $\Delta \log\Xi_{\rm NLR}\sim 1.5$. 
In the detailed
calculations of \cite{groves2006}, the ratio of \nii/\ha is sensitive to
the metallicity if the ionisation parameter is within a range of
$\Delta\log\Xi\sim 2.5$. In the case of the highest sensitivity to the ionisation
parameters, we find \nii/\ha$\propto \Xi^{0.3-0.5}$ (for a fixed metallicity of
4$ Z_{\odot}$). The ionisation parameters of the 
present sample are within the range of  validity of the Groves et al. models. 
This justifies the use of the \nii/\halpha line ratios as 
a proper metallicity indicator. The recent grid of state-of-the-art models by
\citet{kewley2013} confirms these trends.
%In comparison, in the calculations of
%\cite{groves2006}, the range of the ionisation parameter is about
% $\Delta\log\Xi_{\rm NLR}\sim 3.0$
%which is much wider than in the present sample.
%This supports that
%\nii/\halpha\ and \nii/\oiii\  are
%good indicators of metallicity of the present sample.

\begin{figure}
\includegraphics[angle=-90,width=\columnwidth]{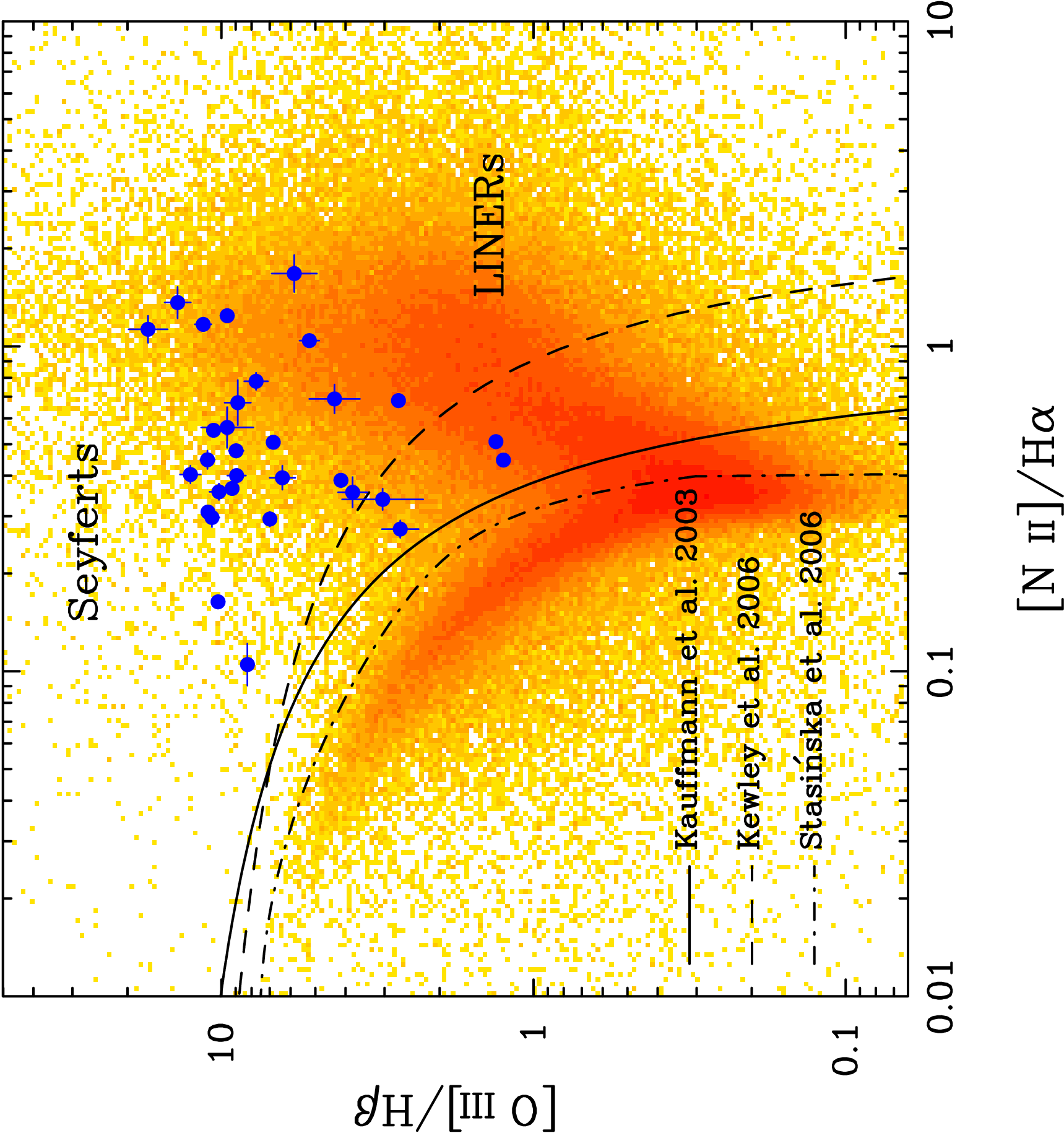}
\caption{The Baldwin-Phillips-Terlevich diagnostic diagram of the present sample of 31
AGNs (blue filled circles). The yellow background is the sample of Seyfert 2, star-forming galaxies, 
H{\sc ii} galaxies, and LINERs from the MPA compilation of the SDSS sample (available
at \url{www.mpa-garching.mpg.de/SDSS}) adopting a $S/N>3$ limit for the
four lines involved. 
The black solid 
\citep{kauffmann2003}, dashed \citep{kewley2006} and dot-dashed lines
\citep{stasinskaetal2006} are different criteria used to
separate AGNs from star-forming galaxies. 
}%
\label{fig4}
\end{figure}

\subsection{Metallicity in the NLR}
\label{NLRmetal}
Figure~\ref{fig4} shows the BPT diagram of the present sample. 
\cite{groves2006} found that, approximately,  
$Z_{\rm NLR}\approx 1.0 + 3.0\left\{\log{\rm [NII]/H\alpha+1}\right\}$
for \nii/H$\alpha\ge 0.1$ (from 
their Figure 2, 
 neglecting the dependence on ionisation parameter). 
We have a range of metallicity
$Z_{\rm NLR}\approx (1-4)Z_{\odot}$ (corresponding
to \nii/H$\alpha$=0.1-1.5) in our sample. This is
further confirmed by recent state-of-the-art models 
\citep{kewley2013} which 
also show that the current sample has this range of metallicities,
irrespective of the SEDs and/or the ionisation parameter \citep[see
their Fig. 4, and note that the updated solar abundance is, in that
scale, of $12+\log(\mathrm{O/H}) = 8.69\pm0.05$, ][]{asplund2009}. 
NLRs have
metallicities ranging between the values found in the bulge and in the BLR,
according to
measurements made in a few objects \citep{matsuoka2009}. These preliminary
results also indicate
that the metallicity in the NLR correlates with emission line luminosity
rather than with redshift \citep{matsuoka2009}.

As Fig.~\ref{fig3} shows, the metallicity of discrete cold
clouds in the NLR follows that of the BLR, indicating an origin closely
connected to the BLR. One natural way for this to happen is AGN outflows, 
which may causally connect the BLR and the
NLR. For example, quite a large fraction of \oiii lines have blue-shifted
wings, indicating the importance of
outflows in the NLR \citep{komossa2008}, and more
generally these outflows are linked to the
feedback processes \citep[see the detailed discussion by][]{nesvadba2006}.
Cold clouds may partially involve the hot gas from stellar
winds in the bulge \citep{ho2009}. We represent the metallicity in the
NLR by
\begin{equation}
Z_{\rm NLR} \; = \; f_{\rm bulge} \, Z_{\rm bulge} \, + \, \left( 1 \, - \,
f_{\rm bulge}\right) \, Z_{\rm BLR} \; ,
\end{equation}
where $f_{\rm bulge}=\dot{M}_{\rm bulge}/(\dot{M}_{\rm bulge}+
\dot{M}_{\rm outflow})$, and
$\dot{M}_{\rm bulge}$ is the mass rate of stellar winds from stars
in the bulge. 
The factor $f_{\rm bulge}$ could be significantly smaller
than unity (for some redshifts) and
$Z_{\rm bulge}\ll Z_{\rm BLR}$, leading to
$Z_{\rm NLR} \, \approx \, (1-f_{\rm bulge})Z_{\rm BLR}< Z_{\rm BLR}$, so
that $Z_{\rm BLR} \, \propto \, Z_{\rm NLR}$ is indeed expected.  
The correlation found in Fig.~\ref{fig3} strongly supports this hypothesis.

\begin{figure*}
\begin{minipage}{120mm}
\includegraphics[angle=-90,width=120mm]{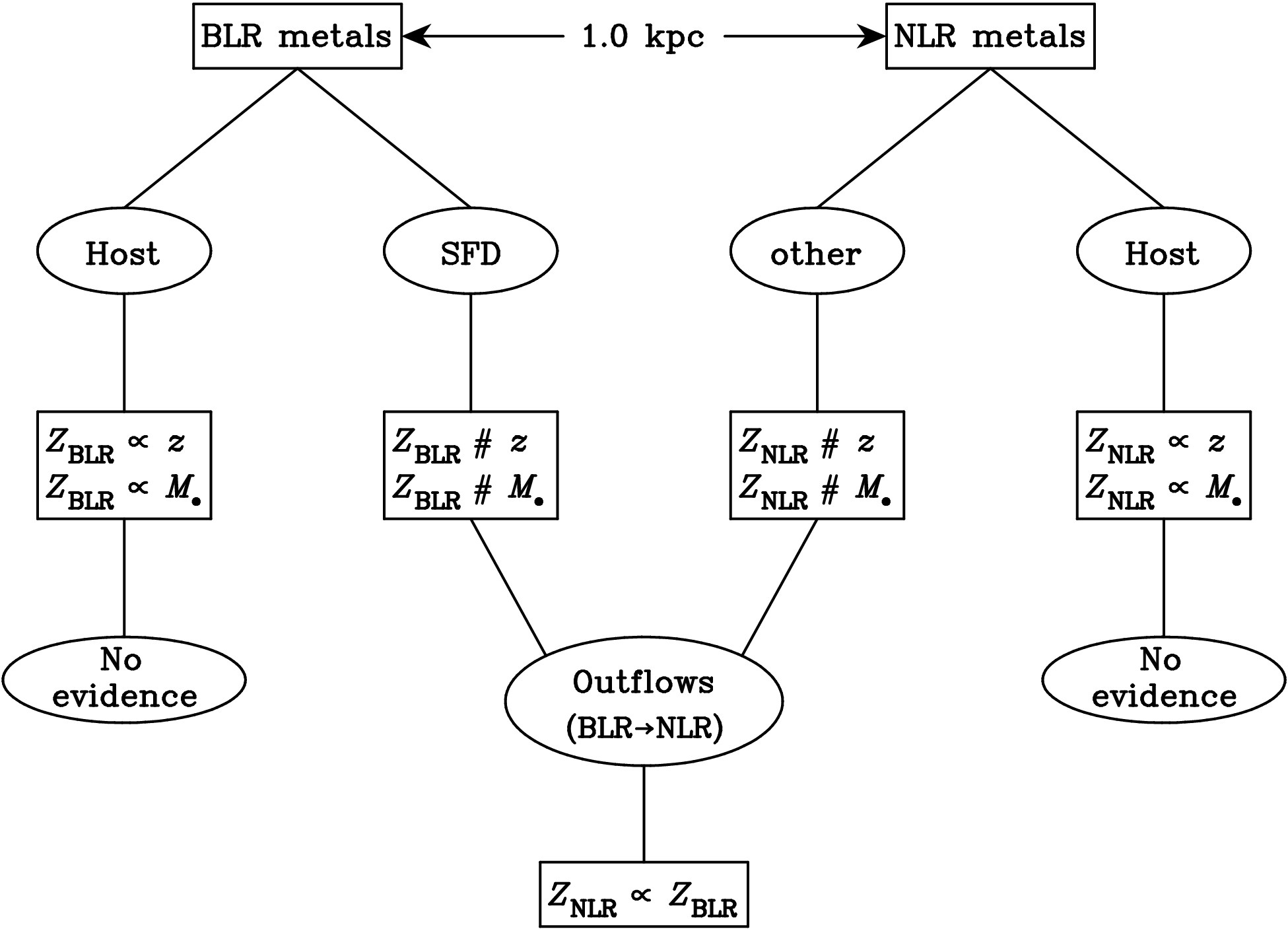}
\caption{Flow chart of the arguments in this paper. If metals are
produced by the processes
shown in the ovals on the second line, correlations
should follow as indicated on the third line. The results found
in this work favour that BLRs and NLRs are connected through
outflows developed from BLRs. ``SFD" refers to
``star forming disc", which is the self-gravity-dominated
regions of accretion disks \citep[e.g.][]{wang2012}.
``Other" refers to any processes for NLR metal production
except for the normal chemical evolution of the host
galaxy, including star formation related to the AGN accretion
process. The symbols $\propto$ and $\#$ indicate
``correlated with" and ``not correlated with", respectively.}
\label{fig5}
\end{minipage}
\end{figure*}

The metallicity in the bulges, for example, of massive star 
forming galaxies ($10^{10-11} \sunm$) is in the range  
$Z_{\rm SF}\approx (2-3)Z_\odot$ \citep[e.g. see Fig. 6 in][]{tremonti2004}, 
whereas the metallicity of NLRs in typical AGNs 
 is of $Z_{\rm NLR}\approx (2-4)Z_\odot$
\citep{kewley2013}.
However, the comparison of $Z_{\rm bulge}$ with $Z_{\rm NLR}$ is open 
for the same
individual objects, leading to
difficulties to estimate the factor $f_{\rm bulge}$. Future
studies on this issue will examine several important physical processes
(see a brief discussion at the end of this section).

A caveat may arise from the fact that while our sample clearly follows 
the expected distribution of Seyfert 2 galaxies with $Z_{\rm NLR} \approx 
(1-4)Z_{\odot}$ in the BPT diagram, it perhaps 
does not trace the underlying density given by the SDSS sample (i.e. 
perhaps a larger number of blue points should lie towards the LINER area in
Fig.~\ref{fig4}). The reason for this effect is unclear, as our sample
is not biased in any of the quantities involved here. It may reflect
the small size of the current sample or perhaps
be linked to selection effects within the framework of AGN unification 
schemes \citep[e.g.][]{zhang2008}.

The chemical abundance connection between BLRs and NLRs found here provides a
new clue for understanding the structure of AGNs. As a brief summary,
Fig.~\ref{fig5} shows a flow chart of the arguments presented in this paper. If
the metals in BLRs are produced by the host galaxies, two correlations are then
expected: $Z_{\rm BLR}\propto z$ and $Z_{\rm BLR}\propto \mbh$. Neither is
supported by any observational evidence. The same reasoning applies to NLRs.
Hence, the only scenario that can explain the present results includes: (1) a
star-forming disc which produces the metals within the BLR clouds; and (2)
outflows produced in the BLR transport gas, including metals, to the NLR.

These outflows will impact the host galaxies. The kinematical effects on the
host galaxy's ISM and on the further growth of the host galaxy are discussed by
\cite{nesvadba2006}.  However, the  BLR-NLR metallicity correlation found here
is a new aspect of AGN feedback, which has additional observable consequences
which we discuss next.

The metallicities of NLRs and of the host galaxies need to be compared for
individual objects in the future.  But we would like to point that the $Z_{\rm
BLR} - Z_{\rm NLR}$ correlation found in this paper indicate that $Z_{\rm NLR}$
should be larger that of the host galaxy, otherwise the correlation should disappear. 
Observations with Integral Field Units (IFU) may explore the
differences in NLR metallicities from their hosts by comparing the metallicity
within the ionisation cone (i.e. the NLR) and outside of the cone. This will
test the chemical evolution and circulation from the BLR and NLR to the host
galaxies.

\subsection{Post-AGN star formation}
The metallicity correlation between BLRs and NLRs has an interesting
implication for the further evolution
of the AGN and its host galaxy. During the active episodic phase, metals
are transported by outflows enriching
the NLRs. The diffusion time scale of metals is
$t_{\rm diff}\sim R/c_s\approx 10^7~R_{\rm 1kpc}T_6^{-1/2}$ years,
where $c_s\approx 10^7T_6^{1/2}{\rm cm~s^{-1}}$ is the sound speed, and
is much longer than the
time scale for metal transportation by outflows
$t_{\rm outflow}\sim R/V_{\rm out}\sim 3.3\times
10^5~R_{\rm 1kpc}V_{0.01}^{-1}$ years, where
$V_{0.01}=V_{\rm out}/0.01c$ (see a brief review by Cappi et al. 2013). This means that metals remain in the
NLR gas clouds
once they have been transported there. The metals will significantly
shorten the cooling timescale\footnote{
Approximated by $t_{\rm cooling}\approx 2.3\times 10^5 \,
n_{\rm ISM,1}^{-1} T_6^{1+\gamma}Z_{0.1}^{-1}$ years,
where $n_{_{\rm ISM,1}}=n_{_{\rm ISM}}/1.0{\rm cm^{-3}}$ is
the ISM density
and $Z_{0.1}=Z_{\rm ISM}/0.1\zsun$ is ISM metallicity.
If $t_{\rm heating}>t_{\rm cooling}$,
the ISM cooling is not balanced by the heating,
and the ISM will condense and form stars. The dissipation time scale
of the kinetic energy of outflows
can be approximated by $t_{\rm outflow}$, and thus
$t_{\rm heating}\approx t_{\rm cooling}$ is expected.}
of the interstellar medium (ISM) in the bulge. Normally
the cooling could be balanced by the AGN heating (radiative and kinetic
of outflows) and other sources of heating. However, as a result of
the metal enrichment of the ISM,
$t_{\rm cooling}$ could be shorter than $t_{\rm heating}$, leading to
star formation as soon as AGNs have faded away
and turned into LINERs (e.g. Wang \& Zhang 2007; Ho 2008). In this case, the accretion rates of black holes
become so low that no broad components
of the emission lines appear in these LINERs, but significant star
formation will begin after
the outflows from the nuclei have been quenched. We will call these
objects {\em blue} LINERs whereas LINERs
without significant star formation will be called {\em red} LINERs.

Detailed calculations are needed in order to fully predict the observable
signatures of this post-AGN star formation in blue LINERs, but the major
features are expected to be: (1) only narrow emission lines should be seen; and
(2) a population of young stars should produce significant near infrared
emission and/or an ultraviolet excess from massive stars. LINERs from blue to
red represent a sequence of decreasing star formation rates, and thus
constitute an evolutionary sequence for the fading of AGNs.  The
fact that the  \nii/\halpha ratio in LINERs is larger than in  star
forming galaxies could imply that the emission line regions in LINERs are
more metal rich than those in star forming galaxies (This should be 
demonstrated
by both theory and observations in the future). This might be caused by
post-AGN star formation further enhancing the metallicity in the LINERs. Future
systematic examinations using {\em Spitzer} and {\em Herschel} observations
should be able to discover these blue LINERs, providing  new clues for the
understanding of the physical processes of AGN fading.

\section{Conclusions}
\label{section:5}
We have carefully assembled in this paper a sample of AGNs and quasars
with redshifts $z \lesssim 0.35$
for which the metallicity of both the BLR and the NLR can be measured
in the optical and UV bands. This
allows us to explore the physical connections between broad and
narrow line regions and we find no correlation
between metallicity and black hole masses or Eddington ratios in either
BLRs or NLRs. However,
we do find that the metallicities in the NLR follow the
metallicities in the BLR, with both being substantially
enriched above the level of the stellar populations in the bulges 
of the host galaxies,
This suggests that metals are produced in the BLRs and then
transported to the NLRs through  outflows.
The metal-enriched NLRs may undergo
star formation once AGNs begin turning into LINERs, that is, some
post-AGN star formation might be triggered.
There is preliminary evidence for this process, but future surveys of
post-AGN LINERs are needed to fully unveil this mechanim.

\section*{Acknowledgements} 
The authors acknowledge useful
comments and suggestions by the anonymous referee. We also 
appreciate the stimulating discussions carried out with members
of the AGN group at IHEP.  This research is supported by grants NSFC-11173023,
-11233003 and the 973 project (2009CB824800). JMW was also supported by the
Sino-French LIA ORIGINS (CNRS/CAS) and by the Observatoire de Paris (poste
rouge).  DVG was partly supported by a visiting professorship for senior
international scientists (Chinese Academy of Sciences).  JAB acknowledges
support from NSF grant AST-1006593.

\appendix
\section{Emission line fittings}
\begin{figure*}
\begin{minipage}{160mm}
\includegraphics[width=160mm]{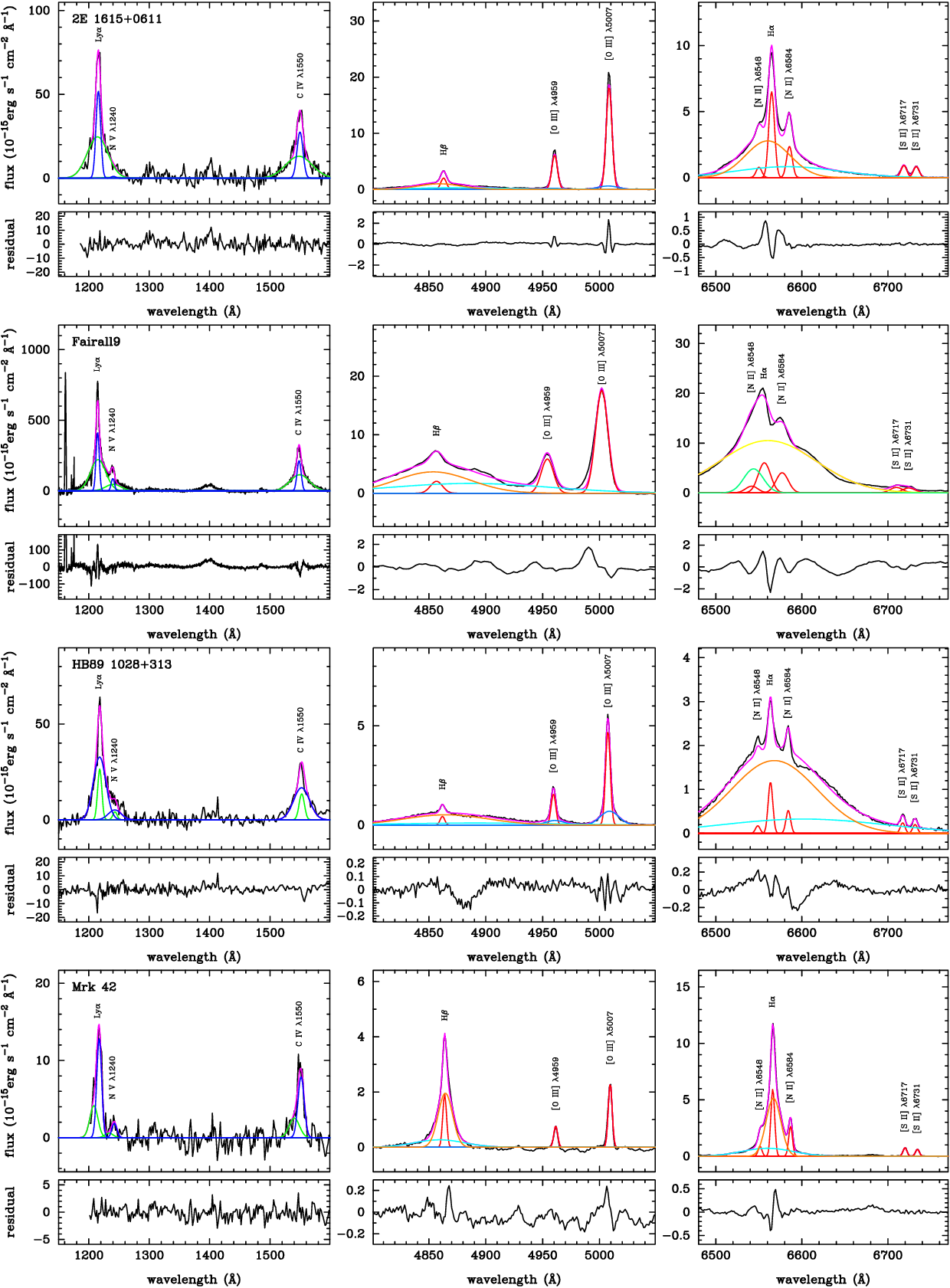}
\caption{Continuum-subtracted spectra and results of 
emission-line fittings.
{\em Left}: UV spectra for \nv/\civ; {\em middle}: optical band
for H$\beta$ and \oiii; and {\em right}: optical band for
H$\alpha$ and \nii. The profiles with
the same colour are constrained to have the same FWHM and shifts. The
magenta curves are the sum of the different components. 
The blue and green lines represent the two Gaussian components of
the broad emission lines of \lya, \nv\, and \civ. The red lines are 
the narrow components
used to calculate the narrow-line ratios, e.g. \oiii/\hb and \nii/\ha.
The light blue lines are the wing components of \oiii\dl4959, 5007. The 
brown and cyan lines are the
broad components of \ha and \hb. The light green and yellow lines 
are the third components
in seven objects whose broad components of \ha and \hb clearly have 
different shapes.  
See the text for details on the fitting procedure (\S 2.3).
Residuals are obtained by subtracting the sum of all the components from 
the spectra, and the unit is $10^{-15}\mathrm{erg~s^{-1}~cm^{-2}~\AA^{-1}}$.}
\label{fig:fits}
\end{minipage}
\end{figure*}

Figure~\ref{fig:fits} presents the details of the fitted spectra
for the individual objects of the present sample and is provided online.
We would like to point out that they are continuum-subtracted spectra,
i.e. subtracting a power-law, Fe {\sc ii} and Balmer continuum.

\begin{figure*}
\begin{minipage}{160mm}
\includegraphics[width=160mm]{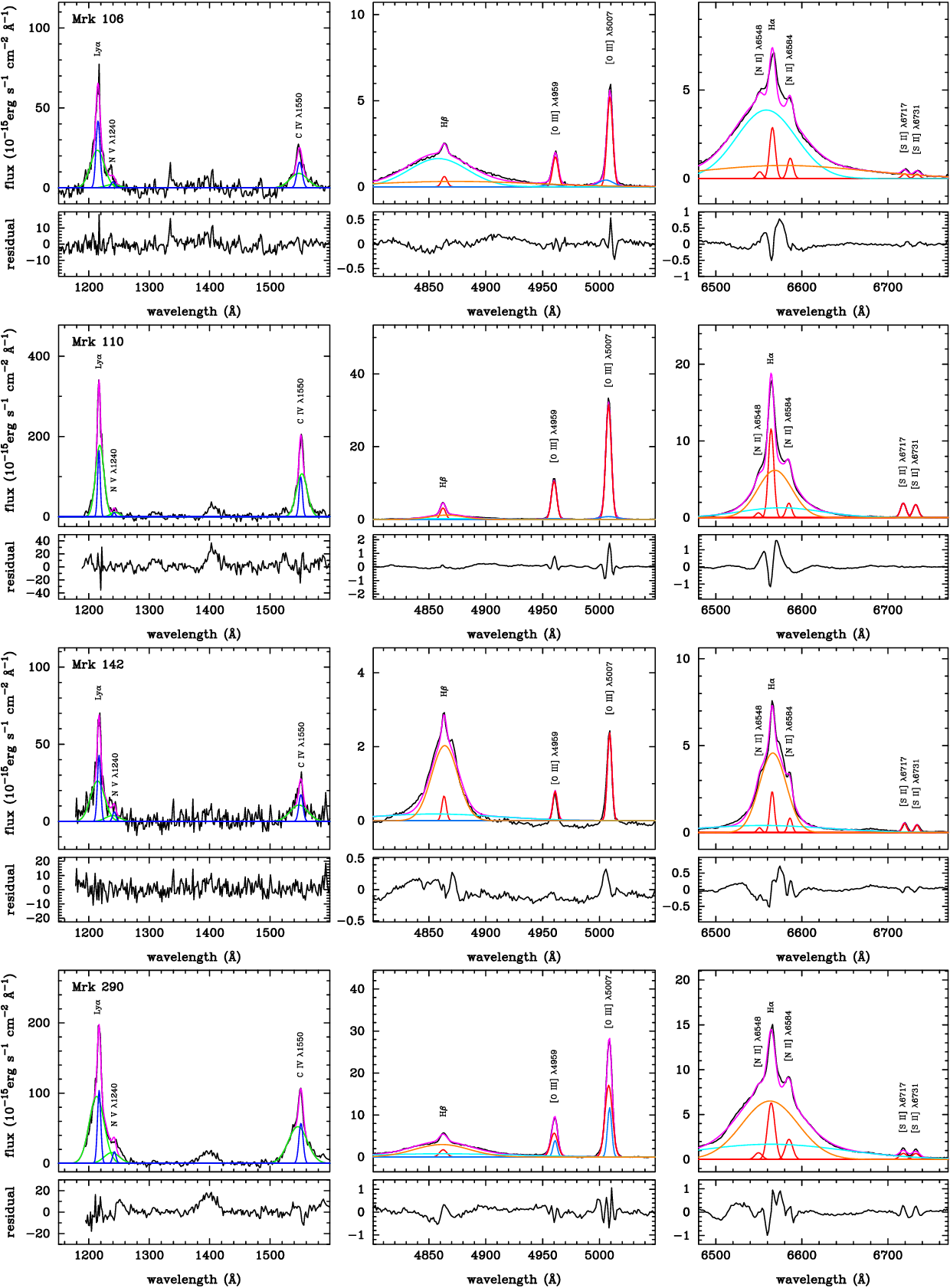}
\caption{Fig.~\ref{fig:fits} continued.}
\end{minipage}
\end{figure*}

\begin{figure*}
\begin{minipage}{160mm}
\includegraphics[width=160mm]{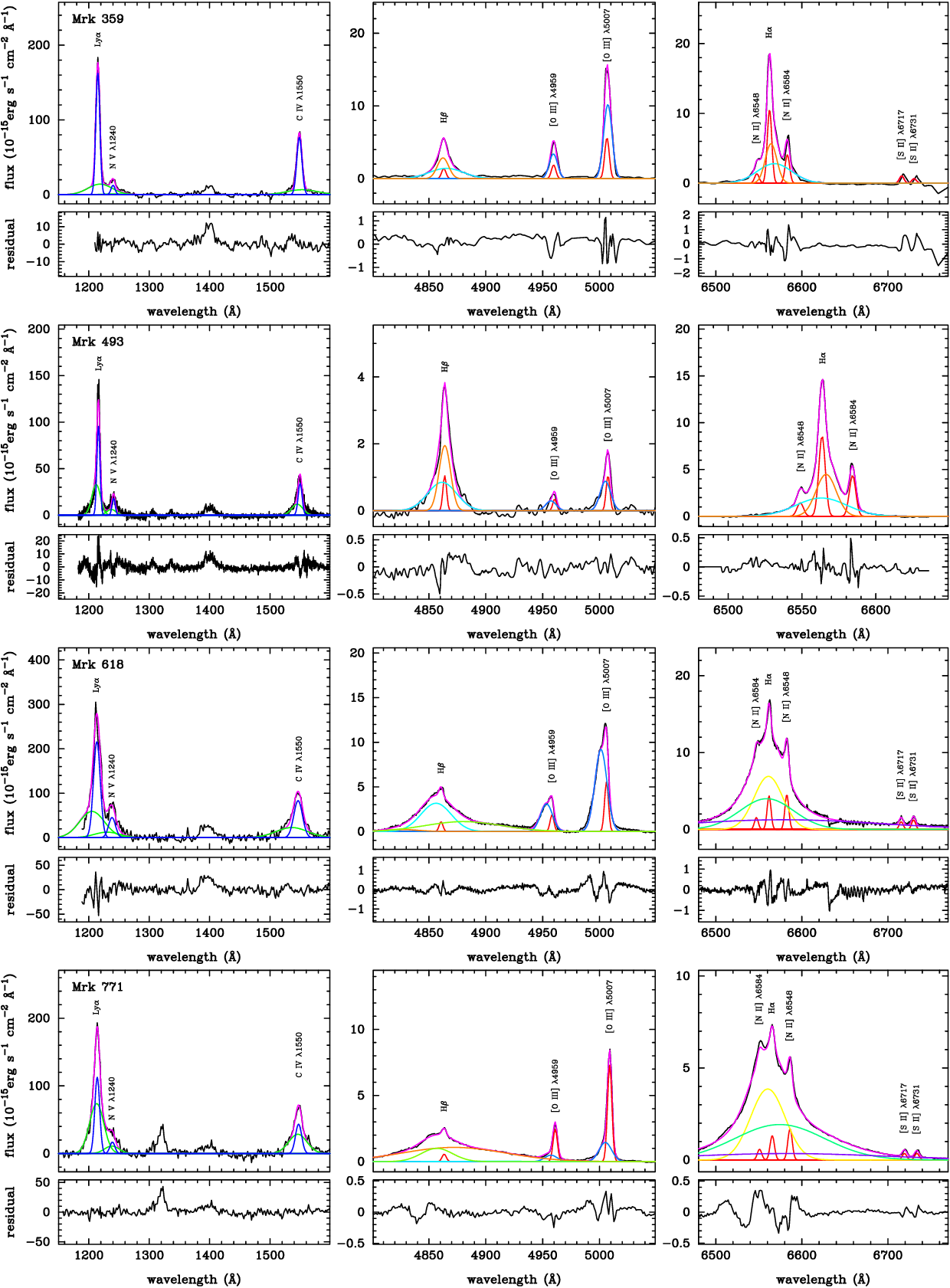}
\caption{Fig.~\ref{fig:fits} continued.}
\end{minipage}
\end{figure*}

\begin{figure*}
\begin{minipage}{160mm}
\includegraphics[width=160mm]{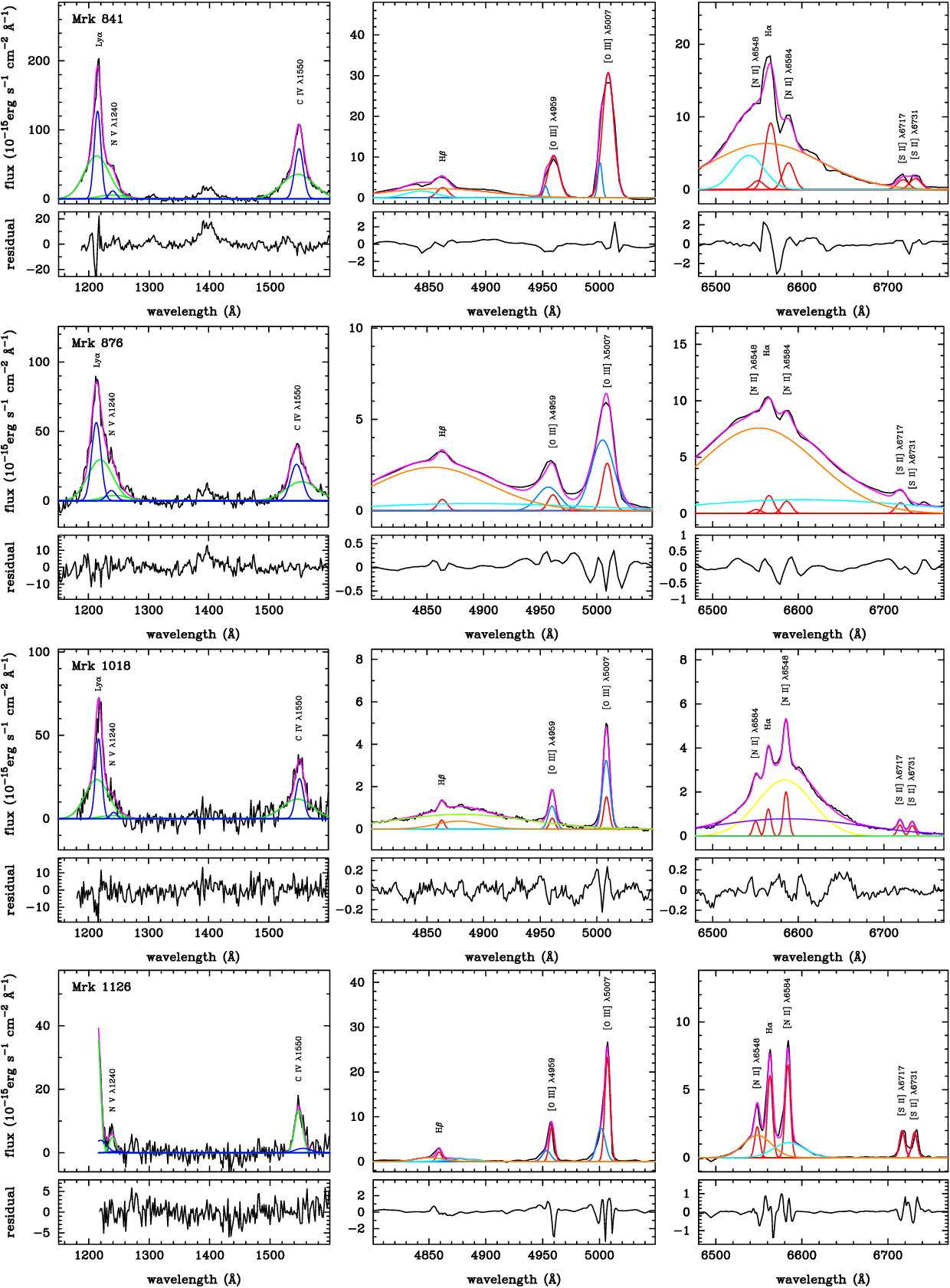}
\caption{Fig.~\ref{fig:fits} continued.}
\end{minipage}
\end{figure*}

\begin{figure*}
\begin{minipage}{160mm}
\includegraphics[width=160mm]{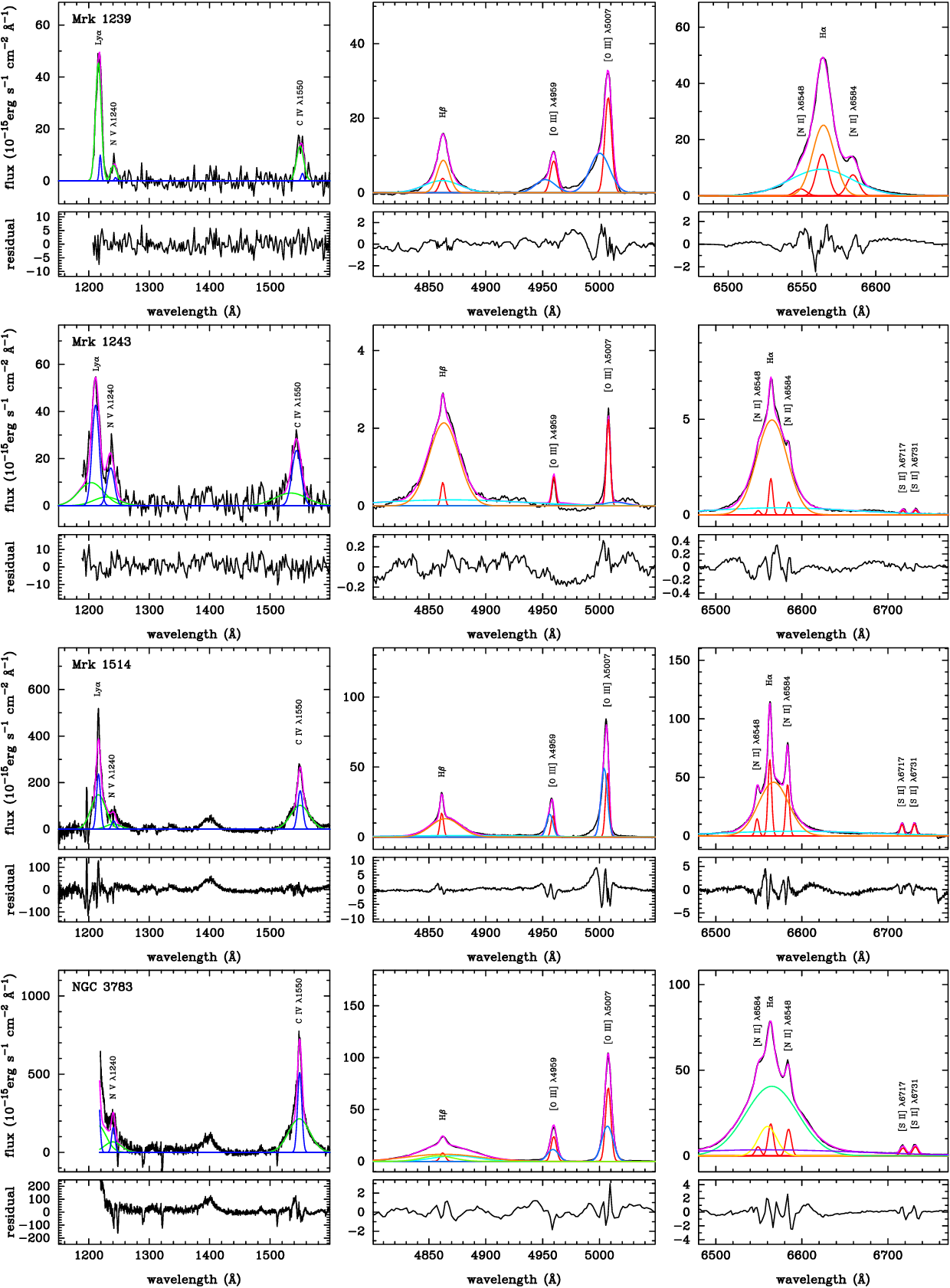}
\caption{Fig.~\ref{fig:fits} continued.}
\end{minipage}
\end{figure*}

\begin{figure*}
\begin{minipage}{160mm}
\includegraphics[width=160mm]{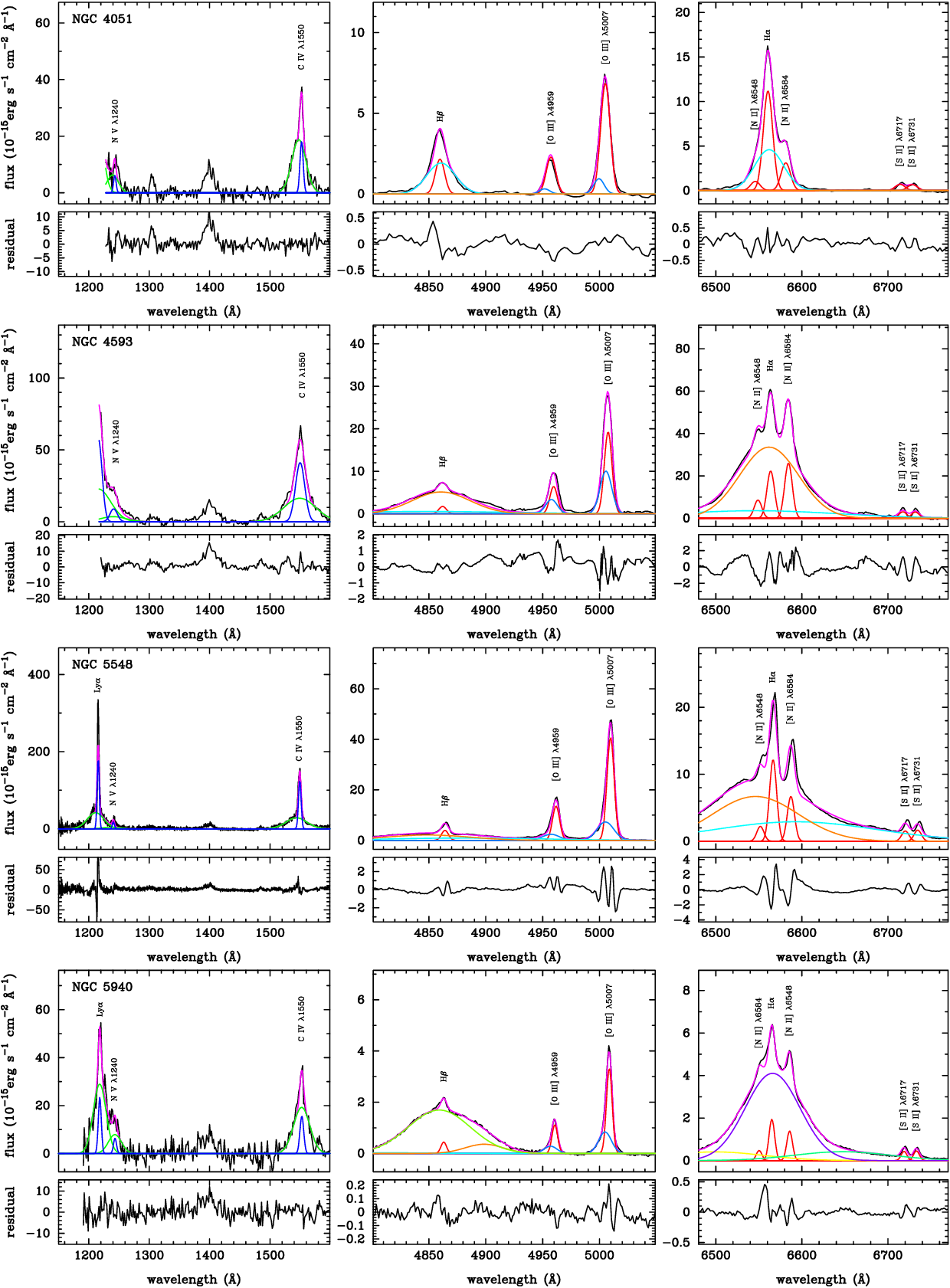}
\caption{Fig.~\ref{fig:fits} continued.}
\end{minipage}
\end{figure*}

\begin{figure*}
\begin{minipage}{160mm}
\includegraphics[width=160mm]{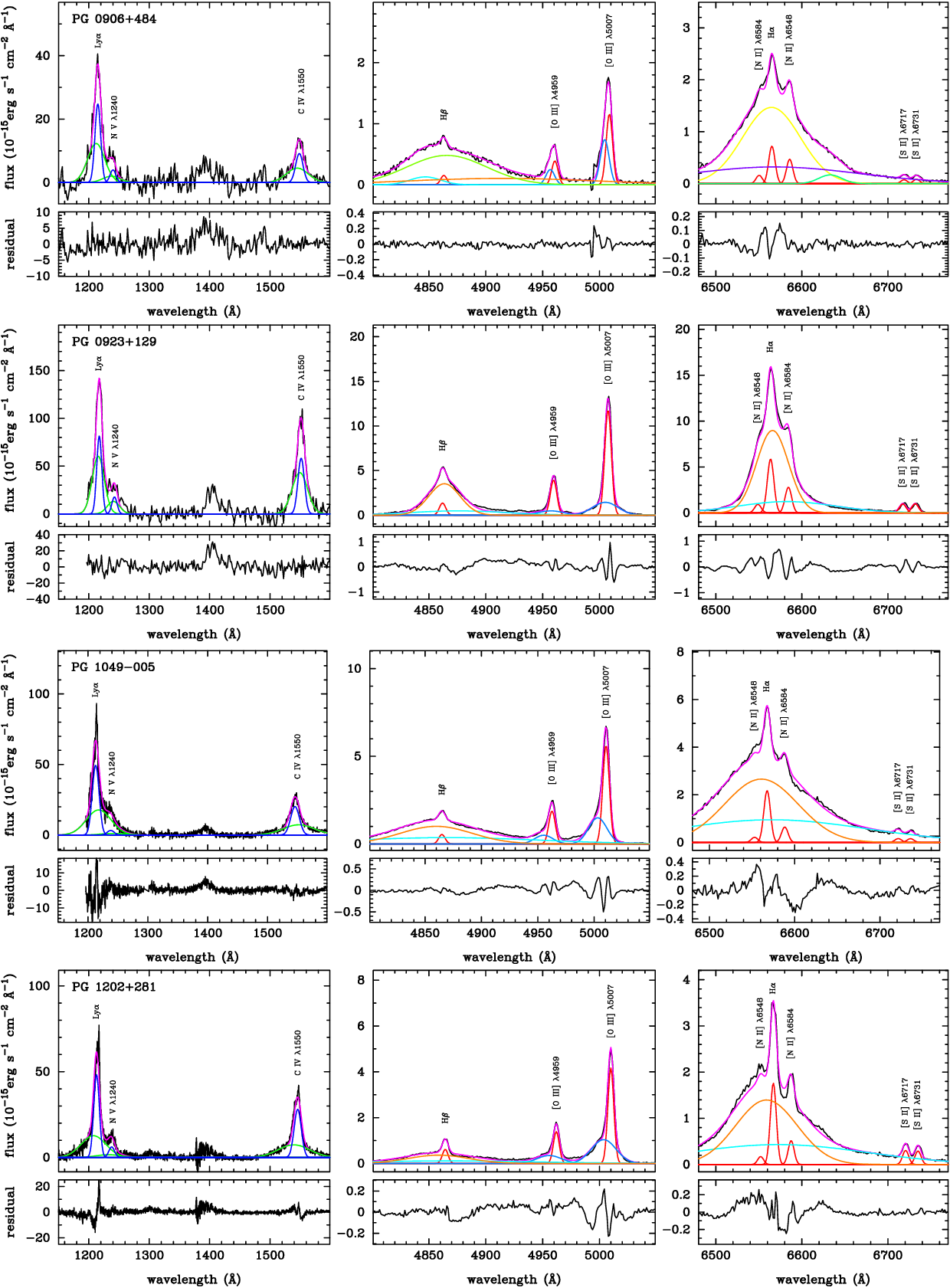}
\caption{Fig.~\ref{fig:fits} continued.}
\end{minipage}
\end{figure*}

\begin{figure*}
\begin{minipage}{160mm}
\includegraphics[width=160mm]{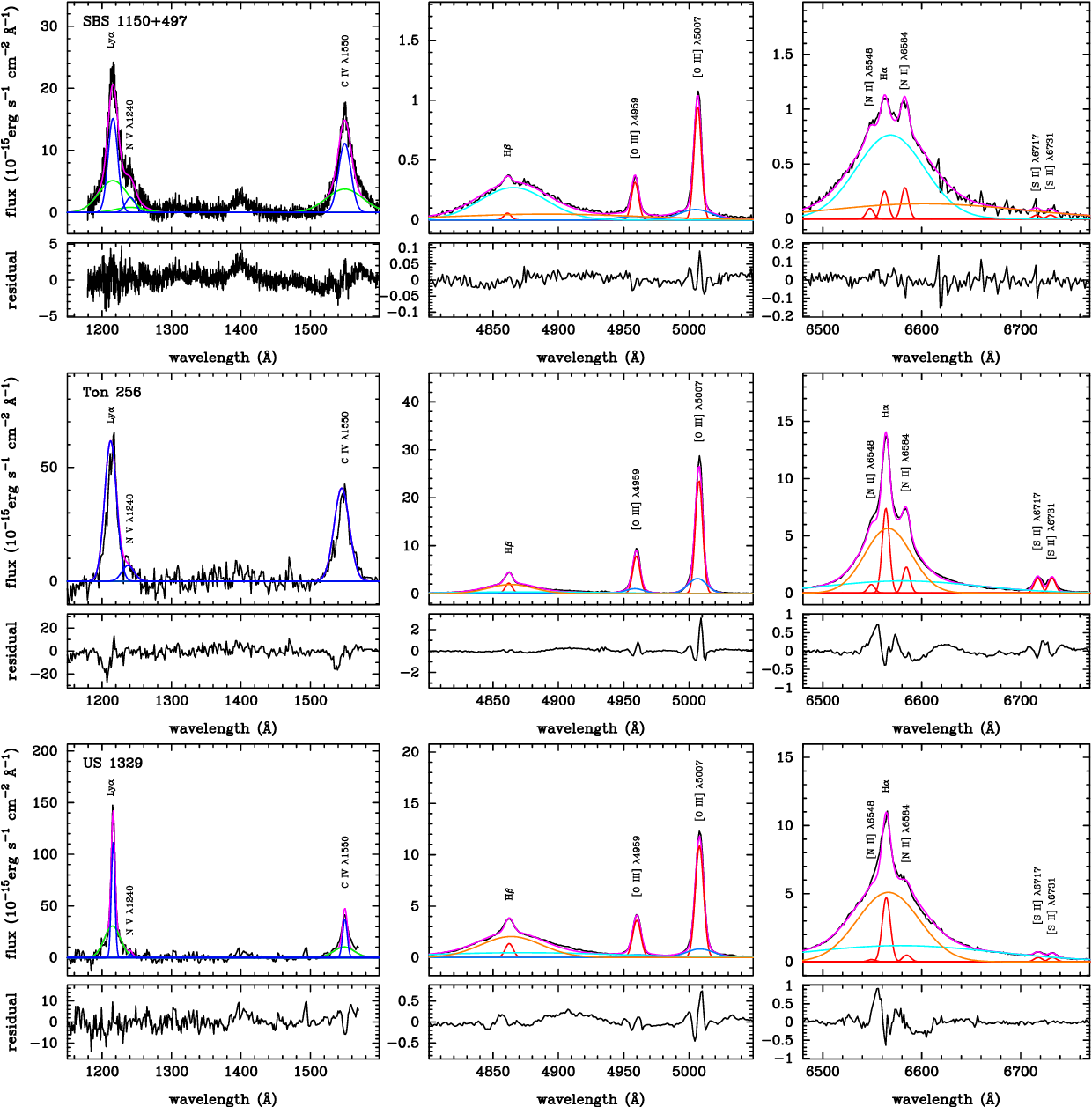}
\caption{Fig.~\ref{fig:fits} continued.}
\end{minipage}
\end{figure*}

\bsp
\label{lastpage}

\begin{thebibliography}{99}
%\bibitem[Alexander \& Hickox(2012)]{alexander2012} Alexander, D. M. \& Hickox, R.C.
%2012, \newastrev, in press (\arxiv{1112.1949})
\bibitem[Asplund et al.(2009)]{asplund2009} Asplund, M. et al. 2009, \araa, 47, 481
\bibitem[Baldwin, Phillips \& Terlevich(1981)]{bpt1981} Baldwin, J. A., Phillips, M. M. \& Terlevich, R. 1981, \pasp, 93, 5
%\bibitem[Baskin \& Laor(2005)]{baskin2005} Baskin, A. \& Laor, A. 2005, \mnras, 358, 1043
\bibitem[Bennert et al.(2002)]{bennert2002}Bennert, N. et al. 2002, \apj, 574, 105
\bibitem[Bentz et al.(2009)]{bentz2009} Bentz, M. et al. 2009, \apj, 697, 160
%\bibitem[B\"ohringer \& Hensler(1989)]{bohringer1989} B\"ohringer, H. \& Hensler, G. 1989, \aap, 215, 147
\bibitem[Brightman et al.(2013)]{brightman2013}Brightman, M. et al. 2013, \mnras, 433, 2485
\bibitem[Bower et al.(2006)]{bower2006}Bower, R. G. et al. 2006, \mnras, 370, 645
%\bibitem[Cao(2007)]{cao2007} Cao, X. 2007, \apj, 659, 950
\bibitem[cappi et al. (2013)]{cappi2013}Cappi, M., Tombesi, F. \& Giustini, M. arXiv1301.7199
%\bibitem[Chokshi \& Turner(1992)]{chokshi1992} Chokshi, A. \& Turner, E. L. 1992, \mnras, 259, 421
\bibitem[Collin \& Zahn(1999)]{collin1999}Collin, S. \& Zahn, J.-P. 1999, \aap, 344, 433
\bibitem[Collin \& Zahn(2008)]{collin2008}Collin, S. \& Zahn, J.-P. 2008, \aap, 477, 419
\bibitem[Croton et al.(2006)]{croton2006}Croton, D. J. et al. 2006, \mnras, 365, 11
\bibitem[Denney et al.(2013)]{denney2013}Denney, K.D., Pogge, R.W., Assef, R.J. et al. 2013, \apj, 775, 60
\bibitem[Dhanda et al.(2007)]{dhanda2007} Dhanda, N., Baldwin, J. A., Bentz, M. C., \&  Osmer P. S. 2007, \apj, 658, 804
\bibitem[Dietrich et al.(1994)]{dietrich1994} Dietrich, M. et al. 1994, \aap, 284, 33
\bibitem[Dietrich et al.(2002)]{dietrich2002} Dietrich, M. et al. 2002, \apj, 581, 912
\bibitem[Di Matteo et al.(2005)]{dimatteo2005}Di Matteo, T., Springel, V. \& Hernquist, L. 2005, \nat, 433, 604
\bibitem[Dopita et al.(2000)]{dopita2000} Dopita, M.A. et al. 2000, \apj, 542, 224
%\bibitem[Dors et al.(2008)]{dors2008} Dors, O. L., Storchi-Bergmann, T., Rirffel, R. A. \& Schmidt, A. A. 2008, \aap, 482, 59
\bibitem[Erb et al. (2006)]{erb2006}Erb, D. K. et al. 2006, \apj, 644, 813
%\bibitem[Fall \& Rees(1985)]{fall1985}Fall, S. M. \& Rees, M. J. 1985, \apj, 298, 18
%\bibitem[Ferland et al.(1998)]{ferland1998} Ferland, G. J., et al. 1998, \pasp, 110, 761
\bibitem[Fu \& Stockton(2007)]{fu2007} Fu, H. \& Stockton, A. 2007, \apj, 664, L75
\bibitem[Giannuzzo \& Stirpe(1996)]{giannuzzo1996} Giannuzzo, M. E. \& Stirpe, G. M. 1996, \aap, 314, 419
\bibitem[Greene et al.(2011)]{greene2011} Greene, J. et al. 2011, \apj, 732, 9
\bibitem[Groves, Dopita \& Sutherland(2004)]{groves2004} Groves, B. A., Dopita, M. A. \& Sutherland, R. S. 2004, \apjs, 153, 75
\bibitem[Groves, Heckman \& Kauffmann(2006)]{groves2006} Groves, B. A., Heckman, T. M. \& Kauffmann, G. 2006, \mnras, 371, 1559
\bibitem[Grupe et al.(2004)]{grupe2004} Grupe, D. et al. 2004, \aj, 127, 156
%\bibitem[Haiman \& Menou(2000)]{haiman2000} Haiman, Z. \& Menou, K. 2000, \apj, 531, 42
\bibitem[Hamann \& Ferland(1992)]{hamann1992} Hamann, F. \& Ferland, G. 1992, \apj, 391, L53
\bibitem[Hamann \& Ferland(1993)]{hamann1993} Hamann, F. \& Ferland, G. 1993, \apj, 418, 11
\bibitem[Hamann \& Ferland(1999)]{hamann1999} Hamann, F. \& Ferland, G. 1999, \araa, 37, 487
\bibitem[Hamann et al.(2002)]{hamann2002} Hamann, F. et al. 2002, \apj, 564, 592
%\bibitem[Hao et al.(2005)]{hao2005} Hao, L. et al. 2005, \aj, 129, 1783
%\bibitem[Heckman et al.(2004)]{heckman2004} Heckman, T. et al. 2004, \apj, 613, 109
\bibitem[Ho(2008)]{ho2008} Ho, L. C. 2008, \araa, 46, 475
\bibitem[Ho(2009)]{ho2009} Ho, L. C. 2009, \apj, 699, 638
%\bibitem[Ho et al.(2009)]{hoetal2009} Ho, L. C. et al. 2009, \apjs, 183, 1
\bibitem[Hu et al.(2008)]{hu2008} Hu, C. et al. 2008, \apj, 687, 78
%\bibitem[Isobe et al.(1990)]{isobe1990} Isobe, T., Feigelson, E. D., Akritas, M. G., Babu, G. J. 1990, \apj, 364, 104
\bibitem[Juarez et al.(2009)]{juarez2009}Juarez, Y., Maiolino, R. \& Nagao, T. 2009, \aap, 494, L25
\bibitem[Jiang et al.(2008)]{jiang2008}Jiang, L., Fan, X. \& Vestergaard, M. 2008, \apj, 679, 962
\bibitem[Kaspi et al.(2000)]{kaspi2000} Kaspi, S. et al. 2000, \apj, 533, 631
%\bibitem[Kauffmann \& Haehnelt(2000)]{kauffmann2000} Kauffmann, G. \& Haehnelt, M. 2000, \mnras, 311, 576
\bibitem[Kauffmann et al.(2003)]{kauffmann2003} Kauffmann, G. et al. 2003, \mnras, 346, 1055
%\bibitem[Kelly \& Merloni(2012)]{kelly2012} Kelly, B.C., Merloni, A. 2012, \advast, in press (\arxiv{1112.1430})
\bibitem[Kewley \& Dopita(2002)]{kewley2002}Kewley, L. J. \& Dopita, M. A. 2002, \apjs, 142, 35
\bibitem[Kewley et al.(2006)]{kewley2006} Kewley, L., Groves, B., Kauffmann, G. \& Heckman, T. 2006, \mnras, 372, 961
\bibitem[Kewley et al.(2013)]{kewley2013} Kewley, L.J., Dopita, M.A., Leitherer, C. et al. 2013, \apj, 774, 100
%\bibitem[King \& Pringle(2006)]{king2006} King, A. R. \& Pringle, J. E. 2006, \mnras, 373, L90
\bibitem[Kollatschny, Zetzl \& Dietrich(2006)]{kollatschny2006} Kollatschny, W., Zetzl, M. \& Dietrich, M. 2006, \aap, 454, 459
%\bibitem[Kollmeier et al.(2006)]{kollmeier2006} Kollmeier, J. A. et al. 2006, \apj, 648, 128
\bibitem[Komossa et al.(2008)]{komossa2008} Komossa, S., Xu, D., Zhou, H., Storchi-Bergmann, T. \& Binette, L. 2008,
\apj, 680, 926
%\bibitem[Korista et al.(1996)]{korista1996} Korista, K., Hamann, F., Gerguson, J. \& Ferland, G. 1996, \apj, 461, 641
\bibitem[Kuraszkiewicz et al.(2002)]{kuraszkiewicz2002} Kuraszkiewicz, J. K. et al. 2002, \apjs, 143, 257
\bibitem[Kuraszkiewicz et al.(2004)]{kuraszkiewicz2004} Kuraszkiewicz, J. K. et al. 2004, \apjs, 150, 165
%\bibitem[Li, Wang, \& Ho(2008)]{li2012} Li, Y.-R., Wang, J.-M. \& Ho, L. C. 2012, \apj, 749, 187
\bibitem[Magorrian et al.(1998)]{magorrian1998} Magorrian, J. et al. 1998, \aj, 115, 2285
\bibitem[Marconi et al.(2004)]{marconi2004} Marconi, A. et al. 2004, \mnras, 351, 169
\bibitem[Matsuoka et al.(2009)]{matsuoka2009} Matsuoka, K. et al. 2009, \aap, 503, 721
\bibitem[Matsuoka et al.(2011a)]{matsuoka2011a} Matsuoka, K. et al. 2011a, \aap, 527, 100
\bibitem[Matsuoka et al.(2011b)]{matsuoka2011b} Matsuoka, K. et al. 2011b, \aap, 532, L10
\bibitem[Mortlock et al.(2012)]{mortlock2012}Mortlock, D. J. et al. 2012 \nat, 474, 616
\bibitem[Moustakas \& Kennicutt(2006)]{moustakas2006} Moustakas, J. \& Kennicutt, R. 2006, \apjs, 164, 81
\bibitem[Mullaney \& Ward(2008)]{mullaney2008} Mullaney, J. R. \& Ward, M. J. 2008, \mnras, 385, 53
\bibitem[Nagao et al.(2006a)]{nagao2006a}Nagao, T., Maiolino, R. \& Marconi, A. 2006a, \aap, 447, 863
\bibitem[Nagao et al.(2006b)]{nagao2006b}Nagao, T., Maiolino, R. \& Marconi, A. 2006b, \aap, 447, 157
\bibitem[Nagao et al.(2006b)]{nagao2006c}Nagao, T., Maiolino, R. \& Marconi, A. 2006c, \aap, 459, 85
\bibitem[Nesvadba et al.(2006)]{nesvadba2006}Nesvadba, N., Lehnert, M., Eisenhauer, F. et al. 2006, \apj, 650, 693
\bibitem[Netzer  et al.(2004)]{netzer2004} Netzer, H. et al. 2004, \apj, 614, 558
\bibitem[Netzer \& Trakhtenbrot(2007)]{netzer2007} Netzer H. \& Trakhtenbrot, B. 2007, \apj, 654, 754
\bibitem[Onken et al.(2004)]{onken2004} Onken, C. A. et al. 2004, \apj, 615, 645
\bibitem[Park et al.(2012)]{park2012} Park, D. et al. 2012, \apjs, 203, 6
\bibitem[Peterson et al.(2004)]{peterson2004} Peterson, B. M. et al. 2004, \apj, 613, 682
\bibitem[Richstone et al.(1998)]{richstone1998} Richstone, D. et al. 1998, Nature, 395, A14
\bibitem[Santos-Lle\'o et al.(1997)]{santos1997} Santos-Lle\'o, M. et al. 1997, \apjs, 112, 271
%\bibitem[Schawinski et al.(2010)]{schawinski2010} Schawinski, K. et al. 2010, \apj, 724, L30
\bibitem[Schmitt et al.(2003)]{schmitt2003} Schmitt, H. R. et al. 2003, \apj, 597, 768
\bibitem[Shakura \& Sunyaev(1973)]{shakura1973} Shakura, N. I., Sunyaev, R. A. 1973, \aap, 24, 337
\bibitem[Shapley et al. (2005)]{shapley2005} Shapley, A. E., Coil, A. L., Ma, C.-P \& Bundy, K. 2005, \apj, 635, 1006
\bibitem[Simard et al. (2011)]{simard2011}Simard, L. et al. 2011, \apjs, 196, 11
\bibitem[Shemmer \& Netzer(2002)]{shemmer2002}Shemmer, O. \& Netzer, H. 2002, \apj, 567, L19
\bibitem[Shemmer et al.(2004)]{shemmer2004}Shemmer, O. et al. 2004, \apj, 614, 547
\bibitem[Shemmer et al.(2006)]{shemmer2006}Shemmer, O. et al. 2006, \apj, 646, 29
%\bibitem[Shen et al.(2008)]{shen2008} Shen, Y. et al. 2008, \apj, 680, 169
\bibitem[Shin et al.(2013)]{shin2013} Shin, J. et al. 2013, \apj, 763, 58
\bibitem[Smith et al.(2004)]{smith2004} Smith, J. E. et al. 2004, \mnras, 350, 140
%\bibitem[So{\l}tan(1982)]{soltan1982} So{\l}tan, A. 1982, \mnras, 200, 115
%\bibitem[Somerville et al.(2008)]{somerville2008} Somerville, R. et al. 2008, \mnras, 391, 481
\bibitem[Small \& Blandford(1992)]{small1992} Small, T. A. \& Blandford, R. D. 1992, \mnras, 259, 725
\bibitem[Stasi\'nska(2006)]{stasinska2006} Stasi\'nska, G. 2006, \aap, 454, L127        
\bibitem[Stasi\'nska et al.(2006)]{stasinskaetal2006} Stasi\'nska, G., Cid Fernandes, R., Mateus, A., Sodr\'e, L.,
       Asari, N. V. 2006, \mnras, 371, 972
\bibitem[Stirpe et al.(1994)]{stirpe1994} Stirpe, G. M. et al. 1994, \apj, 425, 609
\bibitem[Storchi-Bergmann \& Pastoriza(1989)]{storchi1989} Storchi-Bergmann, T. \& Pastoriza, M. G. 1989, \apj, 347, 195
\bibitem[Storchi-Bergmann(1991)]{storchi1991} Storchi-Bergmann, T. 1991, \mnras, 249, 404
\bibitem[Storchi-Bergmann et al. (1998)]{storchi1998} Storchi-Bergmann, T. et al. 1998, \aj, 115, 909
%\bibitem[Trakhtenbrot et al.(2011)]{trakhtenbroat2011}Trakhtenbroat, B. Netzer, H., Lira, P. \& Shemmer, O. 2011,
%        \apj, 730, 7
%\bibitem[Tremaine et al.(2002)]{tremaine2002}Tremaine, S. et al. 2002, \apj, 574, 740
\bibitem[Tremonti et al.(2004)]{tremonti2004}Tremonti, C. A. et al. 2004, \apj, 613, 898
\bibitem[Verner et al.(2004)]{verner2004}Verner, E. et al. 2004, \apj, 611, 780
\bibitem[V\'eron-Cetty, V\'eron \& Gon\c{c}alves(2001)]{veron2001} V\'eron-Cetty, M.-P., V\'eron, P.
\& Gon\c{c}alves, A. C. 2001, \aap, 372, 730
%\bibitem[Wang et al. (2009)]{wangj2009}Wang, J., Wei, J.-Y. \& Xiao, P. F. 2009, ApJ, 693, L66
\bibitem[Wang, Watarai \& Mineshige(2004)]{wang2004} Wang, J.-M., Watarai, K.-Y. \& Mineshige, S. 2004, \apj, 607, 107
\bibitem[Wang et al.(2006)]{wang2006} Wang, J.-M., Chen, Y.-M. \& Zhang, F. 2006, \apj, 647, L17
\bibitem[Wang \& Zhang (2007)]{wang2007} Wang, J.-M. \& Zhang, E.-P. 2007, \apj, 660, 1072
\bibitem[Wang et al.(2008)]{wang2008} Wang, J.-M., Chen, Y.-M., Yan, C.-S. \& Hu, C. 2008, \apj, 673, L9
%\bibitem[Wang et al.(2009)]{wang2009} Wang, J.-M. et al. 2009, \apj, 697, L141
\bibitem[Wang et al.(2010)]{wang2010} Wang, J.-M. et al. 2010, \apj, 719, L148
\bibitem[Wang et al.(2011)]{wang2011} Wang, J.-M., Ge, J.-Q., Hu, C., Baldwin, J. A., Li, Y.-R., Ferland,
G., Yan, C.-S., Xiang, F. \& Zhang, S. 2011, \apj, 739, 3
\bibitem[Wang et al.(2012)]{wang2012} Wang, J.-M., Du, P., Baldwin, J. A., Ge, J.-Q., Hu, C. \& Ferland, G.
2012, \apj, 746, 137
%\bibitem[Warner et al.(2002)]{warner2002} Warner, C., et al. 2002, \apj, 567, 68
\bibitem[Warner, Hamann \& Dietrich(2003)]{warner2003} Warner, C., Hamann, F. \& Dietrich, M. 2003, \apj, 596, 72
\bibitem[Warner, Hamann \& Dietrich(2004)]{warner2004} Warner, C., Hamann, F. \& Dietrich, M. 2004, \apj, 608, 136
\bibitem[Woo et al.(2010)]{woo2010} Woo, J.-H. et al. 2010, \apj, 716, 269
\bibitem[Woo et al.(2013)]{woo2013} Woo, J.-H. et al. 2013, \apj, 772, 49
%\bibitem[Xu \& Cao(2010)]{xu2010} Xu, Y. \& Cao, X. 2010, \apj, 716, 1423
%\bibitem[Yu \& Tremaine(2002)]{yu2002} Yu, Q. \& Tremaine, S. 2002, \mnras, 335, 965
%\bibitem[Yu, Lu \& Kauffmann(2005)]{yu2005} Yu, Q., Lu, Y. \& Kauffmann, G. 2005, \apj, 634, 901
\bibitem[Zhang et al. (2008)]{zhang2008} Zhang, K., Wang, T.-G., Dong, X.-B. \& Lu, H.-L. 2008, \apj, 685, L109
\end{thebibliography}
\end{document}